\documentclass[fleqn,usenatbib]{mnras}

\usepackage{newtxtext,newtxmath}
\usepackage{colortbl}

\usepackage[T1]{fontenc}

\DeclareRobustCommand{\VAN}[3]{#2}
\let\VANthebibliography\thebibliography
\def\thebibliography{\DeclareRobustCommand{\VAN}[3]{##3}\VANthebibliography}

\usepackage{graphicx}	
\usepackage{amsmath}	
\usepackage{soul}
\usepackage{newtxtext,newtxmath}

\definecolor{cyan}{rgb}{0.63, 0.79, 0.95}
	\definecolor{cyan}{rgb}{0.0, 0.72, 0.92}
	\definecolor{magenta}{rgb}{1.0, 0.72, 0.77}
\definecolor{green}{rgb}{0.5, 1.0, 0.83}
\definecolor{green}{rgb}{0.53, 0.66, 0.42}

\newcommand{\maga}{{\sc GANN}}

\title[ML for GA]{Machine Learning for Galactic Archaeology: A chemistry-based neural network method for identification of accreted disc stars}
\author[T. Tronrud et al.]{
Thorold Tronrud,$^{1}$\thanks{E-mail: thor.tronrud@gmail.com}
Patricia B. Tissera,$^{2,3}$
Facundo A. Gómez$^{4,5}$, Robert J. J. Grand$^{6}$, Ruediger Pakmor$^{7}$
\newauthor{ 
 Federico Marinacci$^{8}$ and Christine M. Simpson$^{9,10}$
}
\\
$^{1}$Department of Astrophysics, Universidad Andres Bello, Fernandez Concha 700, Las Condes, Santiago, Chile.\\
$^{2}$Institute of Astronomy, Pontificia Universidad Cat\'olica de
Chile, Av. Vicu\~n Mackenna 4860, 782-0436 Macul, Santiago, Chile.\\ 
$^{3}$Centro de Astro-Ingenier\'ia, Pontificia Universidad Cat\'olica de
Chile, Av. Vicu\~n Mackenna 4860, 782-0436 Macul, Santiago, Chile.\\
$^{4}$Instituto de Investigaci\'on Multidisciplinar en Ciencia y Tecnolog\'ia, Universidad de La Serena, Ra\'ul Bitr\'an 1305, La Serena, Chile.\\
$^{5}$Departamento de Astronom\'ia, Universidad de La Serena, Av. Juan Cisternas 1200 Norte, La Serena, Chile.\\
$^{6}$Instituto de Astrof\'isica de Canarias, Calle Via Lactea s/n, E-38205 La Laguna, Tenerife, Spain.\\
$^{7}$Max Planck Institut fur Astrophysik, Kark-Schwarzschild-Straße, 1, 85648 Garching Bei München, Germany.\\
$^{8}$Department of Physics \& Astronomy, "Augusto Righi", University of Bologna, via Gobetti 93/2, 40129 Bologna, Italy.\\
$^{9}$Enrico Fermi Institute, The University of Chicago, Chicago, IL 60637, USA.\\
$^{10}$Department of Astronomy \& Astrophysics, The University of Chicago, Chicago, IL 60637, USA.\\
}

\date{Accepted 2022 July 14. Received 2022 July 13; in original form 2022 June 3}

\pubyear{2022}

\begin{document}
\label{firstpage}
\pagerange{\pageref{firstpage}--\pageref{lastpage}}
\maketitle


\begin{abstract}
We develop a method ('Galactic Archaeology Neural Network', \maga) based on neural network models (NNMs) to identify accreted stars in galactic discs by only  their chemical fingerprint and age, using  a suite of simulated galaxies from the Auriga Project. We train the network on the target galaxy's own local environment defined by the stellar halo and the surviving satellites.
We demonstrate that this approach allows the detection of accreted stars that are spatially mixed into the disc. Two performance measures are defined \textemdash recovery fraction of accreted stars, $f_{\rm recov}$ and the probability that a star with a positive (accreted) classification is a true-positive result, $P(TP)$. As the NNM output is akin to an assigned probability ($P_{\rm a}$), we are able to determine positivity based on flexible threshold values that can be adjusted easily to refine the selection of presumed-accreted stars.
We find that ~\maga~identifies accreted disc stars within simulated galaxies, with high $f_{\rm recov}$ and/or high $P(TP)$. We also find that stars in Gaia-Enceladus-Sausage (GES) mass systems are over 50\% recovered by our NNMs in the majority (18/24) of cases. Additionally, nearly every individual source of accreted stars is detected at 10\% or more of its peak stellar mass in the disc. We also demonstrate that a conglomerated NNM, trained on the halo and satellite stars from all of the Auriga galaxies provides the most consistent results, and could prove to be an intriguing future approach as our observational capabilities expand.
\end{abstract}




\section{Introduction}


In the $\Lambda$ Cold Dark Matter ($\Lambda$CDM) cosmological model, galaxies undergo  mergers and interactions over the course of their evolution \citep{Springel2008}. These events affect their properties, such as morphology, star formation activity, and angular momentum distribution, as shown by numerous observational \citep[e.g.][]{lambas2003,patton2005,2020NatAs...4..965R,2021ApJ...908...27G} and numerical works \citep[e.g.][]{barneshernquist1991, tissera2000,perez2006,rupke2010,perez2011,2011Natur.477..301P,2013MNRAS.429..159G,amorisco2017, G_mez_2017, moreno2019}. As a consequence, mergers can also shape the properties of the stellar populations that form the different dynamical components: disc \citep[e.g.][]{2009MNRAS.396L..56M,2009MNRAS.397.1599Q,scannapieco_2009,2012MNRAS.423.3727G,2016MNRAS.456.2779G}, bulge \citep[e.g.][]{tissera2018, Gargiulo_2019} and stellar halos \citep[e.g.][]{2005ApJ...635..931B,zolotov2009,font2011,Tissera_2012, Monachesi_2019}. In particular, the stellar halo is expected to be formed mainly by accreted stars with some contribution of in situ star in the inner regions \citep{tissera_2013,Monachesi_2016,brook2020}. It has been shown that the stellar mass function of the accreted satellites leave an imprint in the chemical properties of the stellar halos so that more massive accretion will contribute with high metallicity stars to the central regions while the outer regions will be dominated by the contribution of less massive satellites populated by low metallicity stars \citep{2010MNRAS.406..744C,tissera_2014, Tissera_2018, Dsouza_bell18, Monachesi_2019, Fattahi_2020}. It is worth noting, however, that massive progenitors can also contribute the most metal-poor stars to the inner regions of the stellar halo \citep[e.g.][]{Deason_2016}.

These studies have also shown that the  stellar halos are expected to be populated primarily by the stellar debris of these mergers, which may take the form of coherent tidal features  \citep{Mart_nez_Delgado_2010, 2021arXiv210506467V,2022arXiv220307675M} or  substructures thoroughly spatially-mixed with the halo but which could be disentangled in phase space \citep[e.g.][]{1999Natur.402...53H}. Observations have revealed in the stellar halo of the Milky Way (MW) both type of contributions, which support the claim  that most of the stellar halo has been formed by satellite accretion, including substantial contribution from the Sagittarius stream \citep[e.g.][]{Bell_2008, koppelman2019}. These interactions may lead stars from accreted substructures such as GES \citep{Helmi_2018,Belokurov_2018} to be embedded in the galactic disc.

One of the goals of galactic archaeology is to use the properties of old stars to reconstruct the history of the MW \citep{Helmi_2020}. Accreted stars represent an opportunity to deepen our understanding of specific, far-reaching merger events, and to study the effect these have had on our galaxy's present-day characteristics. First attempts to isolate stars and star clusters based on their motion \citep[e.g.][]{Roman_1950, Eggen_1962, Searle_1978} has contributed to paradigm-shifting theories on galaxy formation that shape our field to this day. As techniques and methods for detecting stellar debris that originated outside the MW improve, the number of known mergers and interactions with the MW increases, and another tiny gap in our understanding of the Galaxy in which we live is filled. By uncovering ancient mergers, we can further test $\Lambda$CDM, and the predictions it makes about galaxy formation.

Previous efforts to identify accreted stars have resorted to a wide variety of techniques. Six-dimensional phase space can be used to identify stars that might have originated from outside the galaxy and formed streams around the MW \citep[e.g.][]{1999MNRAS.307..495H, 2010MNRAS.408..935G,2012MNRAS.423.3727G}. \cite{Borsato_2019} implemented data mining techniques to this phase space data and recovered five stellar streams, one of which had been previously undiscovered. \cite{Malhan_2018} used the STREAMFINDER algorithm to detect a rich network of streams in the MW's stellar halo, including several new structures.  
Specific phase-space parameters, such as angular momentum and radial action can even be used to isolate stars from a single source \citep{feuillet2021selecting}. Alternatively, groups of stars can be separated based on how similar their abundance patterns are to those expected of dwarf galaxies \cite[e.g.][]{font}. \cite{Mackereth_2018} performed this analysis on stars from the MW stellar halo, and determined that 2/3 of these stars display high orbital eccentricity, and enrichment patterns typical of massive MW dwarf satellites today. Their results imply that the MW's accretion history  of dwarf galaxies ($10^{8.5} \leq M_{\rm sat} \leq 10^9$M$\sun$) might have been atypically active at early times  in comparison to other similar galaxies in their {\sc EAGLE} sample. \cite{Fattahi_2019} and \cite{libeskind_2020} report similar findings in both the Auriga and Hestia simulations, respectively.

However, \cite{Kruijssen_2018} reconstructed the MW assembly history through globular clusters, based on a quantitative comparison with simulated analogues. They determined that the MW has undergone no mergers with a mass ratio above 0.25 since at least $\rm z \approx 4$. They did, however, identify three massive satellite progenitors for the ex-situ globular cluster population. This includes a galaxy dubbed "Kraken", to which they attribute 40\% of ex-situ globular clusters \citep[]{Kruijssen_2020, Callingham_2022}. Other techniques based on Machine Learning algorithms, such as boosted decision trees have been utilized on the scale of individual stars by \cite{Veljanoski_2018} to identify MW halo stars. The model was executed on a selection of halo stars from the Gaia Universe Model Snapshot \citep[GUMS]{Robin_2012}. When full phase space data is available, at uncertainties similar to those of Gaia-DR2, 90\% of a halo stars are recovered with 30\% distance errors. 
While these works have focused primarily on halo stars and stellar streams, they display the variety of techniques that are currently available to scientists attempting to disentangle the formation history of our Galaxy. These results also show that, while there have been enormous advances in the study of the MW assembly, many aspects are still far from being fully understood.

Tidal debris from satellites might not only contribute to the stellar halo spheroid, but also to the primarily-in-situ thin and thick discs \citep{Abadi_2003,2015ApJ...799..184P,2015MNRAS.450.2874R, Bignone_2019, Fattahi_2020}.
While galactic stellar discs are expected to form mainly in situ, we also expect a contribution of accreted stars to be distributed by satellites that orbit the primary galaxy near its disc plane. Using a sample of 26 simulated MW mass-sized galaxies from the Auriga Project, \cite{G_mez_2017} showed that, in one third of the models up to 8\% of disc stars, with a circularity parameter $\epsilon_{\rm J}$ above 0.7, had an accreted origin. These fractions increase with lower circularity thresholds, and can make up over 10\% of the stellar mass contained in the population of stars with a circularity above 0.4. These ex-situ disc stars were found to be primarily contributed by one to three massive mergers. In many cases, one of these  donor satellites contributed more than half of the ex-situ disc mass. As such, the identification of their debris would allow us to constrain massive accretion events that could have played a significant impact on the subsequent evolution of the galaxy, even plausibly leaving behind a local dark matter rotating component. According to simulations, a fraction of the oldest population of disc stars in a galaxy is expected to have formed ex-situ and have been imported by merger events \citep{G_mez_2017}. 
Previous results showed similar trends. Using a set of MW mass-sized galaxies from the Aquarius Project, \cite{Tissera_2012} reported a contribution of up to 15\% accreted stars in the disc. These stars were old, low-metallicity, and alpha-enhanced compared to the in-situ populations. In the case of the simulated MW analogue analyzed by \cite{Bignone_2019} in the EAGLE simulations, which was selected to satisfy several observational traits of the MW,  a massive satellite galaxy resembling a GES event  with $M_{\star} = 3.9 \times 10^{9} M_{\sun}$ impacted the disc. 
These authors reported a contribution of stars from this event to the thick disc representing 1.4\% of its final mass and 0.06\% of the thin disc component at $z=0$.

The goal of this work is to introduce a training method based on neural network architecture that can be applied to classify accreted stars in galactic discs, based only on chemical abundances with no external information. This method can be used both to "clean" stellar disc populations of accreted material to better constrain the process of galaxy formation, and also to isolate the accreted stellar matter for the purposes of galactic archaeology. 
Our method will be based on the chemical abundance information and  stellar ages. Chemical abundance patterns have been shown to be useful to identify contributions from different stellar populations \citep{freeman-blandhawthorn2002}. Our results will be expressed in terms of accreted star particle recovery, and precision, which are adaptations of commonly-used metrics to gauge the performance of NNMs. Neural networks have already been trained to label accreted stars by \cite{Ostdiek_2020}. They  successfully trained their network on simulated kinematic data and applied to observed MW stars from the Gaia DR2 catalog. The training set was built with simulated stars from the FIRE simulations \citep{Hopkins_2018}, and applied to observational data using a technique called Transfer Learning, which allows important selection criteria to be maintained, while adapting the NNM to the details of the new dataset. This approach led to the discovery of the Nyx stream, made up of over 200 stars \citep{Necib_2020, Necib_2020b}.

In this work we consider a selection of simulated galaxies from the Auriga suite, covering a wide variety of stellar discs, all with different fractions of accreted material (shown in Table \ref{tab:networkperf}), and with a variety of  accretion histories \citep{Monachesi_2019}. We define the environment of a simulated galaxy as the system made by its stellar halo and surviving satellites from which the training sets will be defined. The diversity in assembly histories and galaxy environments allow us to test our network and training method more generally, to ensure it can provide useful output in a these situations. Additionally, it permits us to study the situations that lead to poor predictive performance.

We also analyse the performance obtained by using  a training set built from stars in a particular Auriga environment to detect accreted stars in stellar discs of different Auriga systems. \cite{G_mez_2017} have previously explored how the assembly history of galaxies in the Auriga simulations impacts the accreted star particles in the disc, which has motivated our initial galaxy-by-galaxy approach. Additionally, we inspect the performance improvement of incorporating multiple galaxies' environments  into a single training set, to examine the ability of a one-size-fits-all training set to describe each galaxy's unique history.

This paper is organized as follows. In Section \ref{section2}, we describe the Auriga simulations. In Section \ref{section3} we explain the adopted neural networks, our model architecture, the training and evaluation method, and define the two performance metrics. 
Section \ref{section4} encompasses the analysis and main results  on the performance of our method in general, and in several tests to assess the performance by using the defined indicators.
This section also discusses the applicability of our method to the MW, as well as several cases that reflect two different weaknesses of our method. Finally, in Section \ref{section6} we present our conclusions, and the aim of our future work.

\section{The Auriga Simulations}
\label{section2}

The Auriga Project \citep{Grand_2017}  comprises of a set of 30 high resolution zoom-in simulations of late-type galaxies within MW mass-sized halos, within the range of $[10^{12}-2\times 10^{12}]$~M$_{\odot}$.
The initial conditions were selected from a dark-matter-only cosmological simulation of the EAGLE project \citep{Schaye_2014, Crain_2015},  performed in a periodic cubic box of $100 $ Mpc side. The initial conditions are consistent with a $\Lambda {\rm CDM}$ cosmology with parameters $\Omega_{\rm m} = 0.307$, $\Omega_{\rm b} = 0.048$, $\Omega_{\Lambda} = 0.693$, and Hubble constant $H_{0} = 100 h {\rm km^{-1} s^{-1} Mpc}$, and $h = 0.6777$ \citep{planck_collab}. We worked with the Auriga simulations run at level 4, with  dark matter (DM) particle mass  of $\sim 4 \times 10^5$~M$_{\odot}$ and initial baryonic cell mass resolutions  of  $\sim 5 \times 10^4$~M$_{\odot}$. DM halos are identified in the simulations using a friends-of-friends algorithm \citep{davis_1985}, and bound substructures were iteratively detected with a SUBFIND algorithm applied \citep{Springel_2005}.
 
Gas was added to the initial conditions, and its evolution was calculated with the magneto-hydrodynamical code AREPO \citep{Springel_2010}.  A variety of physical processes are followed such as gas cooling and heating, star formation, chemical evolution, the growth of supermassive black holes and Supernova and AGN feedback \citep{Vogelsberger_2013, Marinacci_2013}. 
The inter-stellar medium (ISM) is modeled with a two-phase equation of state from \cite{Springel_2003}. Star formation proceeds stochastically in gas with a density above $n = 0.13 {\rm cm^{-3}}$. 
Star formation and stellar feedback models include phenomenological winds \citep{Marinacci_2013}, and chemical enrichment from SNIa, SNII, and AGB stars, with yield tables from \cite{Thielemann2003} and \cite{Travaglio2004}, \cite{Karakas2010}, and \cite{Portinari1998} respectively. Metals are removed from star forming gas through wind particles, which take $1-\nu_{w}$ of the metal mass of the gas cell that creates it, where $\nu_{w} = 0.6$ is the metal loading parameter. The Auriga simulations track eight non-Hydrogen elements, He, C, O, N, Ne, Mg, Si, and Fe. 
Each star particle (hereafter "star particles" and "stars" will be used interchangeably when referring to simulated data) represents a single stellar population (SSP) with a \cite{Chabrier_2003} initial mass function (IMF).   
The model for baryonic physics has been calibrated to reproduce the stellar mass to halo mass function, the galaxy luminosity function, and the cosmic star formation rate density.


In the Auriga simulations, the chemical  elements are distributed into the interstellar medium by star particles, which inject metals into nearby gas cells as they age. As a simulated galaxy evolves, so do the relative ratios of these elements, as previously-enriched gas forms new stars. After 13.6 Gyr, the resulting abundance distributions encode events which took place during this evolution, the impact of which can be detected today. \cite{Grand_2017b} found that two distinct star formation pathways can lead to conspicuous gaps in radial metallicity distributions, consistent with trends noticed by \cite{Bovy_2016}.

While general trends can be mirrored between the Milky Way and simulated galaxies, the slopes and locations are not reproduced exactly. \cite{Grand_2017b} noted a difference in iso-age metallicity distributions, which they postulate to be due to SNII/SNeIa timing (more detail in \cite{Marinacci_2013}), and the IMF (for which \cite{Gutcke_2018} have studied the impact on total galactic metallicity). On the basis of these previous results, the Auriga simulations provide suitable abundance distributions and galaxy assembly  histories to study the impact of accreted stars in the discs.


In fact, the Auriga simulations have been successfully utilised for predictive purposes in a wide variety of situations. Among them, \cite{Monachesi_2016} used the suite to study the metallicity profile of the stellar halo. \cite{Digby_2019} found that the star formation history (SFH) trends with dwarf stellar mass in Auriga were in good agreement with those of Local Group galaxies.

In this work we focus on  24 of the 30 Auriga galaxies. This subset was selected to not have a close companion at $z=0$, and to contain a clear stellar disc. To this end, we  excluded Auriga halos 1, 8, 11, 25, 29, and 30. Data from each simulated galaxy will be referred to as AuID, where ID is the simulation number. Accreted stars, as determined from the simulation merger trees, will also be referred to as "ex-situ" in this work. They are classified as those that were formed while bound to a different DM halo than that which they occupy at $z=0$. Conversely, stars considered to have been formed in-situ are still bound to the same object or their progenitor in which they were born. The in situ classification encompasses
stars formed from gas brought in by accreted satellites, which will be referred to as 'endo-debris' in this work \citep{Tissera_2012}. Gas from these  satellite may have lower $\alpha$-abundances at a given [Fe/H]  than that expected for the disc, due to their bursty star formation history \citep{tissera_2013}.  Hence star particles of this nature will display different chemical patterns, introducing an extra dimension of variation which should be considered to interpret the results from our method.

\section{Machine Learning approach}
\label{section3}
\subsection{Neural Networks}

A neural network mimics the behaviour of neurons in a brain by transmitting signals between nodes, or neurons. One neuron is connected to several in the layer before it, from which it receives signals, and several in the layer after it, to which it transmits its signal.
A neuron weighs each incoming signal depending on the neuron from which it is originating. All incoming signals are summed, and the total is fed into an activation function to be transmitted as a signal to neurons in the next layer.
We train these NNMs by making adjustments to the weights between neurons, the  building blocks of the NN, until the desired output is reached for each class of input. 

We used the Keras package \citep{chollet2015keras} which implements TensorFlow (https://www.tensorflow.org/about/bib). Our network is composed of six layers. The first is simply a batch normalisation layer \citep{ioffe2015batch} which improves network training rate by reducing internal covariate shift due to possible differences between the distributions of parameters in the training and evaluation data sets. The next four layers are composed of 64, 256, 64, and 32 neurons each, with the Scaled Exponential Linear Unit (SELU) activation function \citep{klambauer_2017}. The properties of this activation function allow our NNMs to converge faster than they would otherwise, as well as removing the risk of both vanishing and exploding gradients. On all hidden layers, we implement LeCun Normal initialisation \citep{LeCun_2000}. The final layer is composed of a single sigmoid neuron, which scales the network output between 0 and 1, and allows us to treat it as a probability of a star particle having been accreted. We tested several configurations with varying numbers and distributions of neurons, including a "flat" variant (i.e. every layer has equal numbers of neurons), an "expanding" variant ( with 32, 64, 128, and 256 neurons), and a "contracting" variant (with 256, 128, 64, and 32 neurons). We found that our "standard" configuration, in addition to the "contracting", retained its average predictive performance at both halved and doubled neuron counts, while the other architectures proved to be less stable.

\begin{figure} 
    \includegraphics[width=\columnwidth]{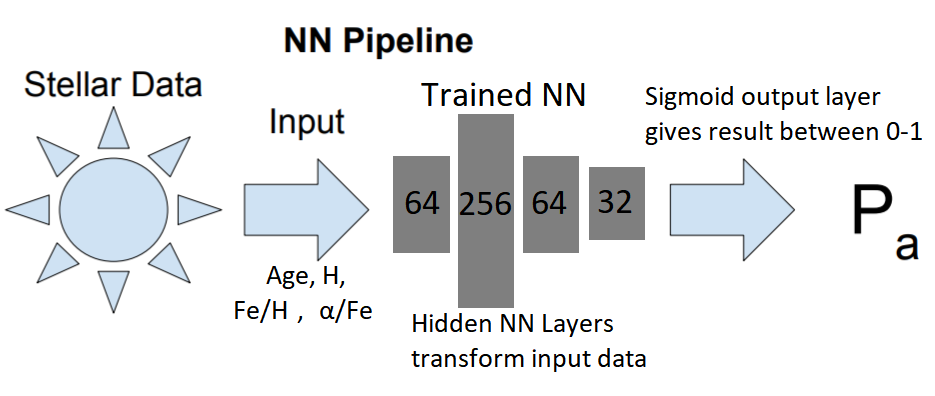}
    \caption{The Neural Network pipeline of \maga. Full star particle data is reduced to  inputs of star particle age (Gyr), hydrogen number fraction, Fe/H and $\alpha$/Fe. The latter two are included as number ratios. These inputs are used by a trained network to assign an accretion probability to the particle. The architecture of the model, and the training method are discussed further in Section \ref{section3}.}
     \label{fig:NNPipeline}
\end{figure}

As our networks are trained with labels between 0 and 1, and our output layer uses a sigmoid neuron, we can interpret the output value for each stellar particle as the probability of being an accreted star, $P_{\rm a}$, as assigned by the network. Star particles that have been assigned a $P_{\rm a}$ by an NNM that is above a certain threshold, $t$ ($t=$ 0.5, 0.75, and 0.9 are used in this work) will be referred to as "positives" or "positively-labelled". Stars assigned a $P_{\rm a}$ by an NNM below 0.5 will be referred to as "negatives" or "negatively-labelled". Hence, accreted stars that are labelled positive are "true-positive", while those labelled negative are "false-negative", which means that they been wrongly classified by the algorithm as in-situ. Similarly, an in-situ star labelled negative is a "true-negative" , while those labelled positive are  "false-positive" and hence, they have been wrongly classified as accreted  (this will be also discussed in Section 3.5).

\subsection{Our Method}

The NNM applied by \cite{Ostdiek_2020}  focused primarily on classifying stars based on parameters that are broadly available across the Gaia DR2 dataset - principally spatial and kinematic data. Interestingly, this work showed that including even the single Fe/H input dimension improved NNM performance in all cases. However, the number of stars with chemical abundance information available at the time of the work limited the applicability of that particular method, and the authors opted to use purely kinematic input. 
Our work is, instead,  entirely based on chemical abundances patterns and age information. The main advantage of this approach is that it does not depend on the degree of phase space mixing of the stellar debris. To this end, we chose to adopt a holistic method of network training using data from the galaxy's environment, i.e, stars in the halo and  surviving satellites, in the same data set as our evaluation and target stars (or in the same suite of observations), avoiding the need for transfer learning entirely \citep{transfer_learning}. 

This also means that any differences between simulated and observed metallicity distributions (as mentioned briefly in Section \ref{section2}) should make little difference to the efficacy of \maga, our description of which includes a specific methodology for constructing training data from the same data source as the population in which one might wish to find accreted stars (see Section \ref{network-training-subsection}).

We have selected H mass fraction ('H'), Fe/H,  and the average of three alpha-iron ratios O/Fe, Mg/Fe, and Si/Fe as our chemical input parameters for ~\maga, and we have opted not to take the logarithm instead leaving them as linear fractions normalized by the solar values. Our figures, however, will use the familiar logarithmic convention.

The logarithm of the latter will  be referred to as [$\alpha$/Fe], following \cite{Bovy_2016}. Additionally, we supply the model with the stellar  age ($\tau$, in Gyrs). While any parameter may be added to the NNM's input, these were chosen based on the primary indicators ([\rm Fe/\rm H] and [$\alpha$/Fe]) used to identify distinct stellar populations that are also tracked in the Auriga simulations \citep{feuillet2021selecting}. 
Including $\tau$ allows the NNM to use accurate ages to differentiate between the diverse temporal evolutionary tracks of metallicity among galaxies of different masses \citep{Venn_2004}. 

Figure \ref{fig:NNPipeline} displays a broad schematic representation of the pipeline through which information will travel in our method. The network, treated as a black box in this depiction, transforms the input parameters into a widely-usable probability value. The specifics of the fully trained network are, however, accessible to the user, who is able to view the weight values in each layer.


\subsection{Network Training}
\label{network-training-subsection}
We constructed our network training sets on a galaxy-by-galaxy basis by creating two sets of stars, labelled as in-situ and ex-situ, with assigned labels 0 and 1, respectively. These sets both contain equal numbers of each classification of star. In a population heavily weighted by an unknown amount towards in-situ stars, this approach prevents the model from de-sensitizing itself to accreted populations. 
The in-situ half of the training set is constructed out of disc stars, of which a small minority are contaminating accreted stars.

We define the disc as the stellar material within $R_{\rm gal} = 0.15 \times R_{200}$, outside the effective bulge radius $R_{\rm eff, b}$ (taken from  \citet{Gargiulo_2019}), with a height along the minor axis of the disc stars between $\pm 10 \rm kpc$, with circularity $\epsilon_J > 0.4$\footnote{This parameter is defined as $\epsilon_J = \frac{L_{z}}{L_{z,\rm max}(E)}$ where  $L_{z}$ is the angular momentum along the main axis of rotation and $L_{z,{\rm max}}(E)$ is its maximum value for all particles of given binding energy $E$}as in \citet{Tissera_2012}. The impact of contamination is expanded on later in this work.

By definition galaxy  satellites are the  "sources" of accreted stars in the galactic disc. In the case of accreted stars present in the primary galaxy's disc at the current epoch, the source may no longer exist. However, in the simulations the source of these stars is identified by the Peak Mass ID, defined as the unique number associated with the most massive object with stellar mass $M_{\rm peak}$ it belonged to prior to becoming bound to the primary galaxy.

The ex-situ half of the training set is sampled from stars beyond 25 kpc from the galactic center, including both halo stars and those belonging to  surviving satellite galaxies.   \cite{Fattahi_2020} have shown that while the innermost regions of the Auriga stellar halo  are dominated by only a few massive progenitors, the outer regions (>20 kpc) have on average 8 main progenitors of relatively lower stellar mass ($M_{\star} < 10^{8} M_{\sun}$) in the Auriga simulations.  Hence, accreted stars typically dominate the mass of the stellar halo beyond 20 kpc \citep[see also][]{Monachesi_2019} as have been previously found in other simulations \citep{zolotov2009, font2011, Tissera_2012}. We have moved our cutoff further to ensure that we do not capture disc stars, and to further minimise contamination by in-situ stars in our halo training sample. Despite this,  both our ex-situ and in-situ training sets may contain some contamination. Although this approach may ignore the contribution of massive satellites to the inner halo, we will sample their abundance distributions directly through surviving satellites, as explained below. 
Even though we are able to separate contaminants from training sets derived from simulated data, we are most interested in realistic applications of \maga, and wish to characterize NNM behaviour in the presence of confounding information, and to demonstrate methods to minimize contamination that do not rely on information unavailable in all circumstances.

\begin{figure} 
    \includegraphics[width=\columnwidth]{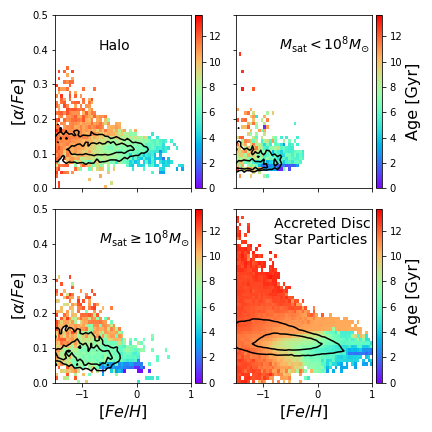}
    \caption{Distributions in [$\alpha$/Fe] and [Fe/H] of the different components of the ex-situ training set for Au14, compared to the true distribution of accreted star particles in the disc (bottom right). While the population sampled from the Halo lie near the true distribution in this space, the sample derived from massive satellites ($M_{\rm sat} \geq 10^{8} M_{\sun}$, bottom left) includes a small knee in low-[Fe/H] space that is absent in the other samples. Black contours enclose 50\% and 90\% of the total mass in each 2D histogram.}
    \label{fig:ex_situ_training}
\end{figure}

For the ex-situ training set, we separate stars bound to surviving satellites by the total stellar mass of the satellite at the current epoch. Very high mass (M$_{\star} > 5 \times 10^{8}$ M$_{\sun}$), high mass ($ 5 \times 10^{8}M_{\odot} >$ M$_{\star} > 10^{8}$ M$_{\odot}$), medium mass ($10^{8}M_{\odot} > M_{\star} > 10^{6} M_{\odot}$), and low mass ($M_{\star} < 10^{6} M_{\odot}$) satellites contribute to the training set as evenly as possible through the imposition of sampling limits in each mass bin. In the event that they contain fewer stars than required, all stars from that source are sampled. This approach  comes with the caveat of underestimating the contribution and the stellar mass of  objects that have been stripped by the central galaxy. However, these stars will likely be part of the outer stellar halo, and thus  considered in the training sets.
As shown in Figure~\ref{fig:ex_situ_training}, these selections, particularly from the halo and massive satellites, span the bulk of the distribution of accreted disc star particles in [$\alpha$/Fe]-[Fe/H] space.

During each cycle of network training, a random sample of the training set is passed through the network. The loss function, a value with which performance is quantified, is determined by comparing the true label values of the training data with those predicted by the network. The aim of training is to minimize this value. In our case, the loss function is binary cross-entropy, which is well-suited for classification problems such as this\footnote{The computation of the binary cross-entropy loss between a true and assigned label can be written as $L\rm(y,z) = \max(z,0) - zy + \log(1 - e^{-\lvert z \rvert})$ where y and z are the true and predicted labels, respectively.}. The network weights are adjusted by the optimizer based on the gradient of the trainable values with respect to the loss. The learning rate can be adjusted dynamically if the loss value plateaus between training cycles. Early stopping is also active, and will restore the best-performing NNM weights once training concludes. In our case, the network is trained over a maximum of 25 cycles, which is enough for the NNM's performance to plateau. Further training iterations have failed to significantly impact NNM performance, in our tests.

\begin{table*}
    \caption{Impact of the specific parameters,hydrogen mass fraction, H, iron-to-hydrogen mass fraction, Fe/H, $\alpha$-to-hydrogen mass fraction, $\alpha/$~H and stellar age, $\tau$, on the performance of the trained NNMs in a selection of three simulated galactic discs. The flattened parameter (Param) denotes which specific parameter has been set to zero to assess its impact. Results are quantified by  two adopted metrics, P(TP) and f$_{\rm rec}$ (see Section 3.5).}
    \label{tab:param_weight}
    \begin{tabular}{rrrrrrr}
        \multicolumn{1}{r|}{Au14} & P(TP) &  & \multicolumn{1}{r|}{} & f$_{\rm rec}$ &  &  \\ \hline
        \multicolumn{1}{r|}{Param} & Pa > 0.5 & Pa > 0.75 & \multicolumn{1}{r|}{Pa > 0.9} & Pa > 0.5 & Pa > 0.75 & Pa > 0.9 \\ \hline
        \multicolumn{1}{r|}{None} & 0.69 & 0.8 & \multicolumn{1}{r|}{0.9} & 0.88 & 0.59 & 0.17 \\
        \multicolumn{1}{r|}{H} & 1 & 1 & \multicolumn{1}{r|}{1} & 0 & 0 & 0 \\
        \multicolumn{1}{r|}{Fe/H} & 0.66 & 0.71 & \multicolumn{1}{r|}{0.91} & 0.462 & 0.387 & 0.156 \\
        \multicolumn{1}{r|}{$\alpha$/Fe} & 0.12 & 0.12 & \multicolumn{1}{r|}{0.12} & 1 & 1 & 1 \\
        \multicolumn{1}{r|}{$\tau$} & 0.3 & 0.36 & \multicolumn{1}{r|}{0.46} & 0.994 & 0.984 & 0.964 \\ \hline
         &  &  &  &  &  &  \\
        \multicolumn{1}{r|}{Au20} & P(TP) &  & \multicolumn{1}{r|}{} & f$_{\rm rec}$ &  &  \\ \hline
        \multicolumn{1}{r|}{Param} & Pa > 0.5 & Pa > 0.75 & \multicolumn{1}{r|}{Pa > 0.9} & Pa > 0.5 & Pa > 0.75 & Pa > 0.9 \\ \hline
        \multicolumn{1}{r|}{None} & 0.75 & 0.84 & \multicolumn{1}{r|}{0.92} & 0.36 & 0.17 & 0.03 \\
        \multicolumn{1}{r|}{H} & 0.78 & 0.78 & \multicolumn{1}{r|}{0.78} & 0 & 0 & 0 \\
        \multicolumn{1}{r|}{Fe/H} & 0.75 & 0.85 & \multicolumn{1}{r|}{0.87} & 0.251 & 0.12 & 0.019 \\
        \multicolumn{1}{r|}{$\alpha$/Fe} & 0.73 & 0.77 & \multicolumn{1}{r|}{0.83} & 0.387 & 0.305 & 0.193 \\
        \multicolumn{1}{r|}{$\tau$} & 0.52 & 0.55 & \multicolumn{1}{r|}{0.56} & 0.703 & 0.608 & 0.497 \\ \hline
         &  &  &  &  &  &  \\
        \multicolumn{1}{r|}{Au22} & P(TP) &  & \multicolumn{1}{r|}{} & f$_{\rm rec}$ &  &  \\ \hline
        \multicolumn{1}{r|}{Param} & Pa > 0.5 & Pa > 0.75 & \multicolumn{1}{r|}{Pa > 0.9} & Pa > 0.5 & Pa > 0.75 & Pa > 0.9 \\ \hline
        \multicolumn{1}{r|}{None} & 0.23 & 0.26 & \multicolumn{1}{r|}{0.33} & 0.99 & 0.97 & 0.91 \\
        \multicolumn{1}{r|}{H} & 1 & 1 & \multicolumn{1}{r|}{1} & 0 & 0 & 0 \\
        \multicolumn{1}{r|}{Fe/H} & 0.29 & 0.32 & \multicolumn{1}{r|}{0.39} & 0.906 & 0.852 & 0.73 \\
        \multicolumn{1}{r|}{$\alpha$/Fe} & 0.23 & 0.27 & \multicolumn{1}{r|}{0.34} & 0.995 & 0.979 & 0.919 \\
        \multicolumn{1}{r|}{$\tau$} & 0.06 & 0.06 & \multicolumn{1}{r|}{0.07} & 1 & 1 & 1 \\ \hline
    \end{tabular}
\end{table*}

\begin{table*}
    \caption{Neural network results with a training set composed of stars from the galactic halo, and those sampled across a variety of satellite masses. For each halo, we present a summary of network performance as the mean $f_{\rm recov}$ and $P(TP)$ of three NNMs separately trained on the same training set data, evaluating the same galactic disc star particles. As NNM training is a stochastic process, each will behave slightly differently. We provide the simulation number, fraction of accreted stars in the disc selection, and the mean recovery fractions, and true-positive rate for  separate runs at three confidence thresholds (columns from left to right).}
    \begin{tabular}{llllllll}
    \hline
    Halo & Disc $f_{\rm acc}$ & $f_{\rm recov} | P_{\rm a} > 0.9$ & $f_{\rm recov} | P_{\rm a} > 0.75$ & $f_{\rm recov} | P_{\rm a} > 0.5$ & $P(TP) | P_{\rm a} > 0.9$ & $P(TP) | P_{\rm a} > 0.75$ & $P(TP) | P_{\rm a} > 0.5$ \\
    \hline
    2 & 0.11 & 0.10 & 0.41 & 0.85 & 0.50 & 0.53 & 0.51 \\
    3 & 0.13 & 0.24 & 0.41 & 0.51 & 0.65 & 0.60 & 0.56 \\
    4 & 0.33 & 0.07 & 0.30 & 0.51 & 0.79 & 0.69 & 0.59 \\
    5 & 0.08 & 0.60 & 0.78 & 0.88 & 0.68 & 0.56 & 0.49 \\
    6 & 0.08 & 0.30 & 0.66 & 0.79 & 0.75 & 0.65 & 0.55 \\
    7 & 0.32 & 0.07 & 0.35 & 0.65 & 0.93 & 0.87 & 0.80 \\
    9 & 0.07 & 0.71 & 0.80 & 0.86 & 0.57 & 0.48 & 0.43 \\
    10 & 0.02 & 0.63 & 0.92 & 0.99 & 0.73 & 0.44 & 0.26 \\
    12 & 0.14 & 0.29 & 0.56 & 0.73 & 0.71 & 0.57 & 0.50 \\
    13 & 0.08 & 0.25 & 0.70 & 0.94 & 0.79 & 0.66 & 0.53 \\
    14 & 0.12 & 0.17 & 0.59 & 0.88 & 0.90 & 0.80 & 0.69 \\
    15 & 0.12 & 0.16 & 0.70 & 1.00 & 0.70 & 0.56 & 0.42 \\
    16 & 0.07 & 0.41 & 0.81 & 0.93 & 0.74 & 0.58 & 0.46 \\
    17 & 0.02 & 0.85 & 0.99 & 1.00 & 0.40 & 0.26 & 0.21 \\
    18 & 0.03 & 0.73 & 0.97 & 1.00 & 0.48 & 0.37 & 0.32 \\
    19 & 0.20 & 0.13 & 0.43 & 0.68 & 0.88 & 0.78 & 0.67 \\
    20 & 0.36 & 0.03 & 0.17 & 0.36 & 0.92 & 0.84 & 0.75 \\
    21 & 0.15 & 0.16 & 0.41 & 0.57 & 0.74 & 0.67 & 0.58 \\
    22 & 0.02 & 0.91 & 0.97 & 0.99 & 0.33 & 0.26 & 0.23 \\
    23 & 0.09 & 0.25 & 0.56 & 0.71 & 0.87 & 0.72 & 0.63 \\
    24 & 0.11 & 0.21 & 0.61 & 0.85 & 0.59 & 0.55 & 0.52 \\
    26 & 0.14 & 0.30 & 0.50 & 0.62 & 0.77 & 0.70 & 0.67 \\
    27 & 0.10 & 0.13 & 0.47 & 0.66 & 0.62 & 0.62 & 0.55 \\
    28 & 0.22 & 0.14 & 0.31 & 0.42 & 0.88 & 0.78 & 0.72 \\
    \hline
\end{tabular}
\label{tab:networkperf}
\end{table*}

\begin{table*}
    \caption{$P(TP)$ (Columns 1-4) and $f_{\rm recov}$ (Columns 5-8) for $ P_{\rm a} > 0.75$ for trained networks on 10,000 star subsets with fixed fractions of accreted stars, $f_{\rm acc}$. Cell colours: blue represents values above 0.75, green above 0.5, and pink below 0.5.}
    \begin{tabular}{lllllllll}
    \hline
    
    &$P(TP)$ &   & & & $f_{\rm recov}$\\
    \textbackslash{}$f_{\rm acc}$ & 0.1 & 0.05 & 0.025 & 0.01 & 0.1 & 0.05 & 0.025 & 0.01 \\
    \hline \\
    2 & \cellcolor{green}0.63 & \cellcolor{magenta}0.44 & \cellcolor{magenta}0.25 & \cellcolor{magenta}0.12 & \cellcolor{magenta}0.50 & \cellcolor{magenta}0.50 & \cellcolor{magenta}0.46 & \cellcolor{magenta}0.48 \\
3 & \cellcolor{cyan}0.85 & \cellcolor{cyan}0.77 & \cellcolor{green}0.59 & \cellcolor{magenta}0.30 & \cellcolor{magenta}0.15 & \cellcolor{magenta}0.14 & \cellcolor{magenta}0.18 & \cellcolor{magenta}0.12 \\
4 & \cellcolor{cyan}0.78 & \cellcolor{green}0.71 & \cellcolor{green}0.56 & \cellcolor{magenta}0.28 & \cellcolor{magenta}0.09 & \cellcolor{magenta}0.10 & \cellcolor{magenta}0.13 & \cellcolor{magenta}0.11 \\
5 & \cellcolor{cyan}0.92 & \cellcolor{cyan}0.84 & \cellcolor{green}0.69 & \cellcolor{magenta}0.50 & \cellcolor{magenta}0.30 & \cellcolor{magenta}0.28 & \cellcolor{magenta}0.34 & \cellcolor{magenta}0.29 \\
6 & \cellcolor{cyan}0.89 & \cellcolor{cyan}0.82 & \cellcolor{green}0.61 & \cellcolor{magenta}0.41 & \cellcolor{green}0.51 & \cellcolor{green}0.52 & \cellcolor{magenta}0.46 & \cellcolor{magenta}0.45 \\
7 & \cellcolor{magenta}0.47 & \cellcolor{magenta}0.33 & \cellcolor{magenta}0.18 & \cellcolor{magenta}0.06 & \cellcolor{magenta}0.22 & \cellcolor{magenta}0.25 & \cellcolor{magenta}0.27 & \cellcolor{magenta}0.19 \\
9 & \cellcolor{cyan}0.98 & \cellcolor{cyan}0.93 & \cellcolor{cyan}0.84 & \cellcolor{green}0.74 & \cellcolor{green}0.65 & \cellcolor{green}0.64 & \cellcolor{green}0.66 & \cellcolor{green}0.63 \\
10 & \cellcolor{cyan}0.96 & \cellcolor{cyan}0.90 & \cellcolor{cyan}0.78 & \cellcolor{green}0.65 & \cellcolor{cyan}0.86 & \cellcolor{cyan}0.85 & \cellcolor{cyan}0.83 & \cellcolor{cyan}0.86 \\
12 & \cellcolor{cyan}0.92 & \cellcolor{cyan}0.82 & \cellcolor{green}0.66 & \cellcolor{magenta}0.41 & \cellcolor{magenta}0.38 & \cellcolor{magenta}0.35 & \cellcolor{magenta}0.37 & \cellcolor{magenta}0.34 \\
13 & \cellcolor{cyan}0.80 & \cellcolor{green}0.60 & \cellcolor{magenta}0.43 & \cellcolor{magenta}0.28 & \cellcolor{green}0.51 & \cellcolor{magenta}0.49 & \cellcolor{magenta}0.50 & \cellcolor{green}0.59 \\
14 & \cellcolor{green}0.64 & \cellcolor{magenta}0.43 & \cellcolor{magenta}0.27 & \cellcolor{magenta}0.13 & \cellcolor{green}0.74 & \cellcolor{green}0.74 & \cellcolor{green}0.70 & \cellcolor{green}0.73 \\
15 & \cellcolor{magenta}0.15 & \cellcolor{magenta}0.08 & \cellcolor{magenta}0.04 & \cellcolor{magenta}0.02 & \cellcolor{green}0.72 & \cellcolor{green}0.69 & \cellcolor{green}0.71 & \cellcolor{green}0.69 \\
16 & \cellcolor{cyan}0.78 & \cellcolor{green}0.66 & \cellcolor{magenta}0.45 & \cellcolor{magenta}0.24 & \cellcolor{green}0.58 & \cellcolor{green}0.54 & \cellcolor{green}0.58 & \cellcolor{green}0.53 \\
17 & \cellcolor{green}0.75 & \cellcolor{green}0.60 & \cellcolor{magenta}0.41 & \cellcolor{magenta}0.21 & \cellcolor{cyan}0.98 & \cellcolor{cyan}0.98 & \cellcolor{cyan}0.97 & \cellcolor{cyan}0.97 \\
18 & \cellcolor{cyan}0.87 & \cellcolor{cyan}0.79 & \cellcolor{green}0.61 & \cellcolor{magenta}0.37 & \cellcolor{green}0.59 & \cellcolor{green}0.61 & \cellcolor{green}0.55 & \cellcolor{green}0.56 \\
19 & \cellcolor{cyan}0.93 & \cellcolor{cyan}0.78 & \cellcolor{green}0.69 & \cellcolor{magenta}0.45 & \cellcolor{magenta}0.19 & \cellcolor{magenta}0.18 & \cellcolor{magenta}0.19 & \cellcolor{magenta}0.17 \\
20 & \cellcolor{cyan}0.79 & \cellcolor{green}0.62 & \cellcolor{magenta}0.47 & \cellcolor{magenta}0.22 & \cellcolor{magenta}0.10 & \cellcolor{magenta}0.10 & \cellcolor{magenta}0.08 & \cellcolor{magenta}0.13 \\
21 & \cellcolor{green}0.64 & \cellcolor{magenta}0.48 & \cellcolor{magenta}0.30 & \cellcolor{magenta}0.12 & \cellcolor{magenta}0.26 & \cellcolor{magenta}0.26 & \cellcolor{magenta}0.27 & \cellcolor{magenta}0.25 \\
22 & \cellcolor{cyan}0.98 & \cellcolor{cyan}0.97 & \cellcolor{cyan}0.92 & \cellcolor{cyan}0.83 & \cellcolor{green}0.56 & \cellcolor{green}0.56 & \cellcolor{green}0.52 & \cellcolor{green}0.55 \\
23 & \cellcolor{cyan}0.80 & \cellcolor{green}0.65 & \cellcolor{green}0.50 & \cellcolor{magenta}0.27 & \cellcolor{magenta}0.27 & \cellcolor{magenta}0.24 & \cellcolor{magenta}0.30 & \cellcolor{magenta}0.26 \\
24 & \cellcolor{cyan}0.86 & \cellcolor{green}0.74 & \cellcolor{green}0.70 & \cellcolor{magenta}0.44 & \cellcolor{magenta}0.27 & \cellcolor{magenta}0.27 & \cellcolor{magenta}0.30 & \cellcolor{magenta}0.27 \\
26 & \cellcolor{green}0.67 & \cellcolor{green}0.58 & \cellcolor{magenta}0.26 & \cellcolor{magenta}0.21 & \cellcolor{magenta}0.20 & \cellcolor{magenta}0.24 & \cellcolor{magenta}0.14 & \cellcolor{magenta}0.27 \\
27 & \cellcolor{green}0.63 & \cellcolor{magenta}0.43 & \cellcolor{magenta}0.26 & \cellcolor{magenta}0.10 & \cellcolor{magenta}0.13 & \cellcolor{magenta}0.12 & \cellcolor{magenta}0.12 & \cellcolor{magenta}0.12 \\
28 & \cellcolor{green}0.72 & \cellcolor{green}0.51 & \cellcolor{magenta}0.40 & \cellcolor{magenta}0.26 & \cellcolor{magenta}0.22 & \cellcolor{magenta}0.19 & \cellcolor{magenta}0.24 & \cellcolor{magenta}0.25\\
    \hline
    \end{tabular}
\label{tab:fixed_size_ptp_frecov}
\end{table*}

\subsection{Network Evaluation}
The chemical abundance of a star particle is inherited from the progenitor gas cells from which it was created. 
Hence, the stellar component can determine chemical patterns which will reflect the properties of the interstellar medium from which they formed, the IMF, and the history of assembly as they can be dynamical perturbed and mixed, for example during mergers.
Our goal is to construct a method to identify accreted stars in well mixed and primarily in-situ stellar populations. For that purpose, we need to construct an evaluation data set that reflects the  properties of the galactic disc, a structure within which it can be difficult to distinguish between kinematic perturbations due to merger activity and those due to the distribution of matter in the galaxy itself \citep{Helmi_2005}. This makes our chemistry-based approach  well-suited to the problem at hand. However, in order to accurately evaluate our method, we need to construct a data set as close to our anticipated usage scenario as possible.
To this end, we construct the evaluation datasets from star particles in the disc region of each galaxy, which we will refer to as "disc stars" for the purposes of this work. As we are not seeking to minimize the fraction of accreted disc stars in this data, our disc selection for the evaluation dataset is less stringent than that used to construct our in-situ training set.  Hence, we consider all stars between $R_{\rm gal}$ and $R_{\rm eff,b}$ with a distance from the disc plane of less than 5 kpc.

\subsection{Performance Measures}
To quantify the performance of our neural networks on stellar classification, we define and use two metrics, based on the ground-truth accretion labels from the simulations, which allows us in this case to descriminate between true- and false-positive and negative results. The first performance metric we introduce is the Recovery Fraction ($f_{\rm recov}$) that  is the number of true-positive accreted stars with an assigned $P_{\rm a}$ above a stated threshold, divided by the total number of accreted stars, as determined by the ground truth. This metric is also known as the "recall" of a NN. The second metric, $P(TP)$,  is the probability that a star is a "true-positive" result, which will be defined as the probability that a given star with an assigned probability $P_{\rm a}$ above a stated threshold $t$  is actually an accreted star, and not a false-positive. This metric is also known as the "precision" of a NN. This is calculated by taking the number of accreted stars with $P_{\rm a}$ greater than a given threshold $t$, and dividing this by the total number of stars with $P_{\rm a} > t$. We can now delineate our performance metrics by thresholds in $P_{\rm a}$, to build a robust understanding of how network confidence varies on a galaxy-by-galaxy basis.

These metrics are generally anti-correlated. A high $f_{\rm recov}$ will typically correspond with a lower true-positive probability, and vice versa. Similarly, increasing the $P_{\rm a}$ threshold for results from a single galaxy will normally increase the true-positive probability, and lower the recovery fraction. This implies that different thresholds may be more suitable for different goals, e.g. if one is attempting to remove all accreted stars from a sample, a lower threshold of $t = 0.5$ will give the highest chance of  sanitizing the data. On the other hand, if one is trying to gather information on specifically accreted stars, a higher threshold of $t = 0.9$ will give the best probability of having a clean dataset.

Stars that were formed ex-situ but were assigned a $P_{\rm a}$ value below the given threshold for positivity are considered to be false-negatives, and correspond inversely to  $f_{\rm recov}$. The fractional impact of an incorrectly-assigned false-negative star depends on the total population of accreted stars in the disc. False-negatives generally occur in cases where certain sets of satellite masses (or sources) are missing from the galactic environment, leaving gaps in the training set of ex-situ stars. Typically, this occurs when  high or very high stellar mass  satellites are missing, and may have already been completely devoured by the massive primary galaxy they are bound to, and potentially leaving a gap in the satellite galaxy mass-metallicity relation that our network implicitly constructs. An additional source of false-negatives may be contamination confusion, due to the selection of in-situ training examples potentially containing a significant number of accreted stars. We have chosen not to remove these to mimic a naive observational selection criteria.

Stars that were formed in-situ but have been assigned a $P_{\rm a}$ above the threshold for positivity are considered false-positives. The number of these stars heavily affects the value of $P(TP)$, as a higher number of incorrectly-assigned in-situ stars as accreted will lower our confidence in the NNM's final assessment. The possible presence of in-situ stars formed from accreted gas particles (so-called endo-debris stars), may contribute to this number. Such stars might exhibit similar or slightly more enriched chemical abundances than those of truly accreted stars depending on how rapidly the gas was transformed into stars after falling into the galaxy.


The impact of the adopted parameters on the performance of the trained NNMs is displayed in Table \ref{tab:param_weight}. Over three specific simulated galactic discs, NNMs trained on the complete data set were deprived of information for one specific parameter by setting its value in all star particles to zero. The impact of this histogram flattening is determined by the change in performance when compared with the default results.
We can see that H, i.e. the Hydrogen mass fraction, heavily impacts the behaviour of the NNMs, as would be expected for a measure that is inversely proportional to total metallicity. The removal of the other parameters impacts the NNMs in different ways. This is due to the NNMs being trained individually for each selected galactic disc, leading to potentially multiple approaches that are tailored specifically to the unique training data. [Fe/H] appears to have a mild impact on NNM performance across the board, however in the case where the $\tau$ parameter is removed entirely, we have found that the impact of [Fe/H] flattening rises dramatically. The NNM trained on Au14 data appears to use the [$\alpha$/Fe] parameter to distinguish between in- and ex-situ stars, whereas in Au20 and Au22 this flattened parameter only barely changes the results. The impact of $\tau$ flattening also varies by Auriga system. In both Au14 and Au22, it appears to be pivotal for NNM discrimination, however in Au20 this effect is significantly lessened.

\section{Analysis and Results}
\label{section4}
\begin{figure} 
    \includegraphics[width=\columnwidth]{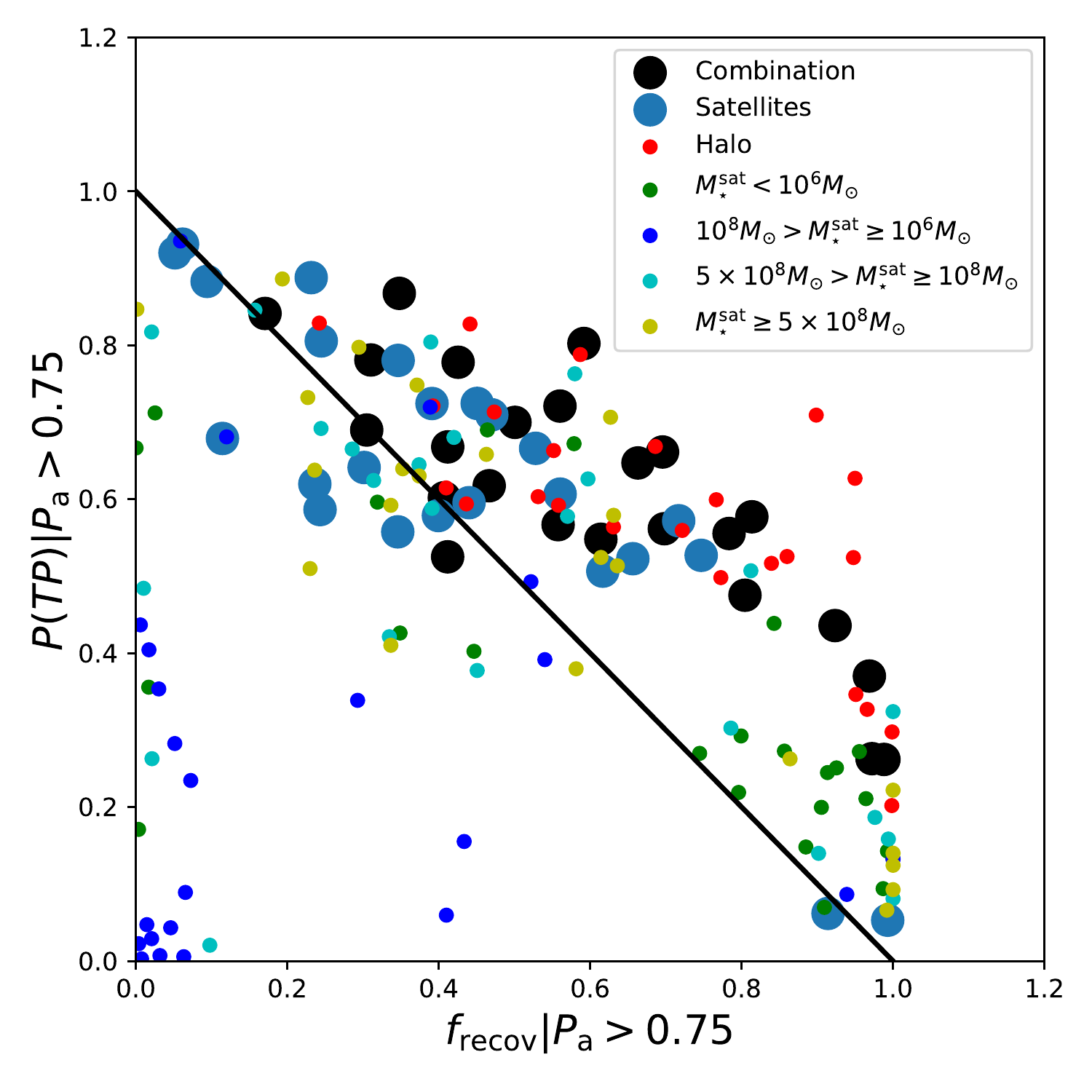}
    \caption{P(TP) as a function of $f_{\rm recov} | P_{\rm a} > 0.75$ for the selected Auriga stellar discs. Symbols are coloured by the specific set of data used to train the NNM, with the final combined set presented as large, black circles.
    The anti-correlation is clear, as is the fact that the NNMs tend to over-perform (e.g. a gain in $f_{\rm recov}$ does not imply an equal loss in $P(TP)$) with respect to the negative 1:1 linear relation (black solid line).}
    \label{fig:ptp_v_frecov}
\end{figure}

In this section, we present the network performance results across the selected Auriga galaxies. 
For this purpose, for each of them, we  constructed the training set by combining stars from the different stellar components, i.e. stellar halo and surviving satellites galaxies as explained in Section 3.2. Then we applied our method to each galaxy.
Table \ref{tab:networkperf} displays the performance indicators of the NNM trained and evaluated on data from each Auriga system. As expected, $f_{\rm recov}$ and $P(TP)$ are anti-correlated. 
In Fig.~\ref{fig:ptp_v_frecov} we display the two metrics for $t = 0.75$ estimated by training over the combined stellar sample (large, black circles). 
First, the anti-correlation is not quite 1:1, with NNMs overperforming in $P(TP)$ as $f_{\rm recov}$ increases. This is good, as it means an increase in one performance measure does not necessarily correspond with an equal decrease in the other. Indeed, our model achieved both P(TP) and $f_{\rm recov}$ above 50\% in 14 of our selected halos. 

Additionally, this figure shows the metrics obtained by training over data sets built from surviving satellites with different stellar mass. As it can be seen,  two individual sets of training data deliver similar NNM performance to the one obtained from the full combined training set: those built from the stellar halo only and from very massive satellites. Specifically, halo stars with galactocentric distances larger than 25 kpc  (red circles) yield generally comparable  P(TP) and $f_{\rm recov}$ at a threshold of $t = 0.75$, although they show a mild systematic increase in $f_{\rm rec}$, and a similar systematic decrease in P(TP). A direct comparison between NNMs trained on halo-only stars and those trained with the combined data set is presented in Fig.\ref{fig:halo_comparison}. We found that, on average, halo-only NNMs (red circles) achieved 9.2\% higher $f_{\rm rec}$ for 2.5\% lower P(TP). In contrast, a training data set composed of satellite stars (large blue circles) evenly split by present-day object mass has on average 2.6\% higher P(TP), and 17.7\% lower $f_{\rm rec}$. 
While surviving satellites, by definition, are expected to have contributed minimally to the accreted population of star particles in the stellar disc, they contain valuable information about age and metallicity which \maga~ can apply during its evaluations to detect star particles that formed in similarly-enriched environments, but whose parent object was destroyed before the current epoch. In particular, massive surviving satellites (small yellow and cyan circles) are relevant to build a good training set. This is important since the inner region of the stellar  halos have been excluded from the training sets. Massive satellites are expected to contribute significantly to these regions \citep[e.g.][]{tissera_2014,Tissera_2018,Fattahi_2020,Khoperskov2022} principally if they are set in radial orbits \citep[e.g.][]{amorisco2017, fernandezalvar2019}.

\begin{figure} 
    \includegraphics[width=\columnwidth]{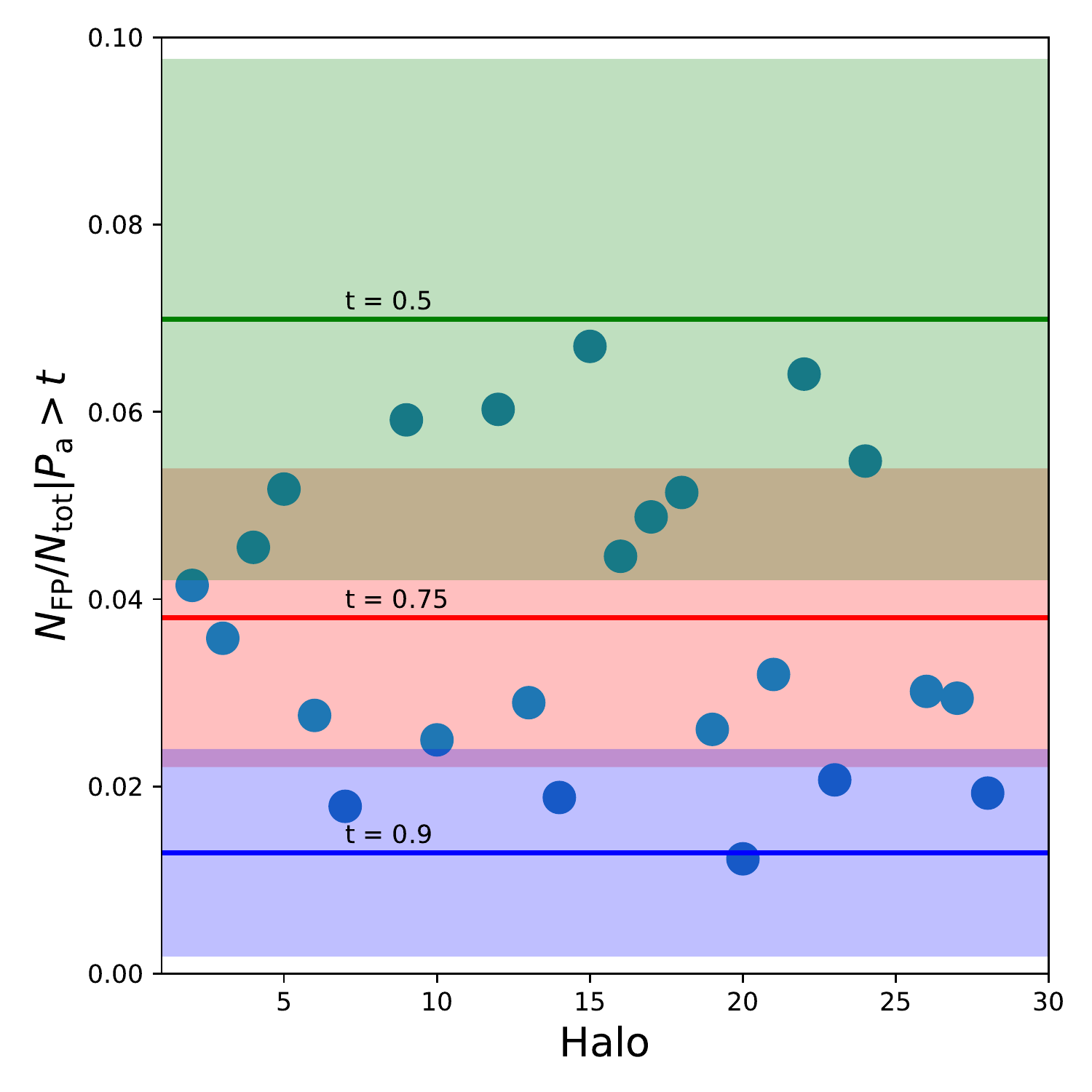}
    \caption{Fraction of the total number of evaluated stars that are false-positives, ordered by their arbitrarily-assigned Auriga  halo number. The circles denote the results for  $t = 0.75$. Mean values for $t = 0.5$ (green), $t = 0.75$ (red), and $t = 0.9$ (blue) are also displayed, with shaded bands corresponding to the standard deviation.}
    \label{fig:median_f_FP}
\end{figure}

Figure \ref{fig:median_f_FP} shows the rate of false-positives in the total evaluation population at a $P_{\rm a}$ threshold of $t = 0.75$ for each Aquarius halo. In situations where we may not know of the exact $P(TP)$ value (e.g. with observational data) this allows us to estimate a median, as well as an upper and lower bound on the number of contaminating false-positives in the end result. For a median false-positivity rate of 4\%, this allows us to calculate: 
$$P(TP)_{\rm prelim} = 1 - 0.04 \times N_{\rm tot} / N(P_{\rm a} > 0.75)$$
without any information besides the total number of stars in the sample. We also include the mean values and the standard deviation for models with $t= 0.9$ and $t =0.5$ for comparison.

To characterize how the network distinguishes between positively- and negatively-labelled material in [$\alpha$/Fe]-[Fe/H] space, we  calculated the median [$\alpha$/Fe] and [Fe/H] of the positive and negative classified star particles, and those of the ground truth accreted and in-situ particles. We find that the median [$\alpha$/Fe] in both positively- and negatively-labelled star particles of 0.13 dex indicates that simply using alpha-element abundance is not enough to differentiate accreted and in-situ star particles at least in the Auriga halos. This could be due to the large scatter in the [$\alpha$/Fe] ratios exhibited by the simulated stellar populations as we will discuss in Section 4.2. The median [Fe/H], however, displays a stark difference. Positively-labelled star particles have a median [Fe/H] of $-0.72$ dex, while negatively-labelled star particles have a median value of [Fe/H]~$= 0.04$ dex. As expected, the ground-truth median [Fe/H], at a value of -0.47 dex across all the Auriga discs, regardless of their origin, is higher than that of the star particles labelled as accreted by the NNMs. This margin decreases to -0.62 dex as the threshold $P_{\rm a}$ decreases to $t = 0.5$.  This implies the NNMs require a larger margin in [Fe/H] than arises naturally to confidently distinguish between the two populations in that space.

We have also found that the fraction of stellar mass with [Fe/H] above and below the galactic disc median value [Fe/H] massively varies between the positively- and negatively-labelled populations of star particles. The fraction of positively-labelled stellar mass that has a [Fe/H] greater than the total galactic disc median is 0.04. For negatively-labelled star particles this fraction is 0.60.

These findings are displayed clearly on the [$\alpha$/Fe]-[Fe/H] plane for for Au14, Au22, and Au20 in the top panels of Fig.\ref{fig:Au14_composite}, \ref{fig:Au22_composite}, and \ref{fig:Au20_composite} which are discussed in detail in the next section. However, to facilitate the interpretation, let us note that the 50\% and 90\% contours only vary slightly in [$\alpha$/Fe] position, while the distinction between the positively and negatively-labelled populations is stark on the [Fe/H] axis.

\subsection{Performance with Fixed Sub-Populations}

To properly evaluate network performance in galaxies with different ex-situ contributions to the stellar disc, we constructed subsamples of stellar disc populations with fixed sample size, while varying the fractions of accreted star particles. Four training subsamples are constructed from the same set of eligible star particles, with fixed total numbers, and fixed fractions of accreted star particles of 10, 5, 2.5 and 1 per cent of a total 10000 star particles. Hence, the difference in performance can be then associated to the different sampling of accreted stellar populations.  Table \ref{tab:fixed_size_ptp_frecov} displays the variation of  $P(TP)$ with accreted fraction. Cells are colour-coded by value \textemdash blue cells are above 0.75, green are above 0.5, and red contain below 0.5. This coding allows us to easily see that $P(TP)$ correlates very well with the fraction of accreted stars in the test subsamples. As the fixed fraction of accreted material is decreased, the rate of false-positives will remain constant, and lead to a decrease in $P(TP)$. 

In summary, Table \ref{tab:fixed_size_ptp_frecov} shows that the recovery fraction of the diminishing accreted population remains relatively constant, which further implies that the cause of the decreased $P(TP)$ is due to the fixed false-positive rate in this experiment. As $f_{\rm recov}$ is a measure of similarity between the accreted populations in the disc and the training satellite populations in the galactic halo, it should not vary significantly as the size of the populations are adjusted, as shown in the table.

\subsection{Accretion Source Recovery}

\begin{figure*} 
    \includegraphics{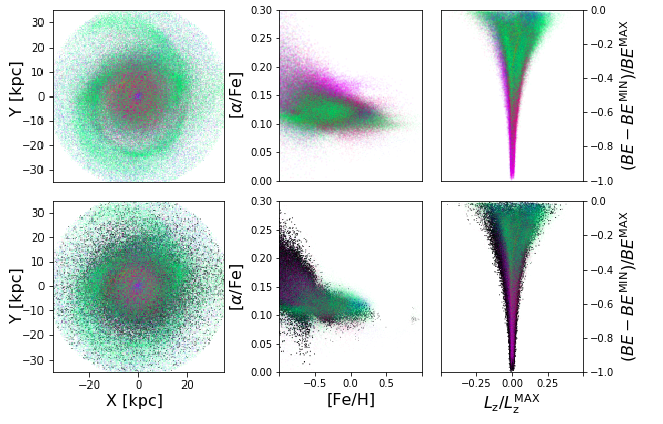}
    \caption{The distribution of ex-situ stars (top), and ex-situ-labelled (bottom) star particles from Au14. Colours represent the star particles brought in by different satellites identified by their unique peak mass Id (see Section 3.3).
    Colours on the bottom panel match those in the top for the same peak mass Id, while black points correspond to mis-labelled in-situ star particles (false-positives). Both the binding energy, BE, and the angular momentum along the z-axis, $L_{\rm z}$, are normalized by the maximum and minimum values to lie in [-1,0], and [-1,1] respectively.}
    \label{fig:h14_source_recovery}
\end{figure*}

The next step is to evaluate the performance of our method to identified accreted stars in the Auriga halos. For this purpose we take advantage of having the complete assembly histories.
Additionally, each accreted stellar particle can be linked to an accreted satellites (or sources) through their Peak mass ID, which is associated to the most massive object the particles were bound to prior to their infall on to the massive, primary galaxy. By comparing these unique source IDs in the accreted stellar particle data sets with $P_{\rm a} > 0.75$ to the total list of unique sources in the discs, we can estimate the power of the NNMs to uncover a picture of the complete accretion history. We find that in every host galaxy in the Auriga set, 85\% or more of the individual accretion sources are recovered at $P_{\rm a} > 0.75$, where we consider a satellite to be recovered if over 10\% of its deposited mass in the present-day stellar disc is correctly labelled by the NNM. This shows the potential of this method to unveil a comprehensive picture of the contribution of accreted stars to the galactic disc, if used in concert with additional  techniques to disentangle the detected accreted stars based on their progenitor satellites. 
As an example, Fig. \ref{fig:h14_source_recovery} displays the spatial distribution and relevant properties of the accreted stars coloured by their sources (top panels), and the populations that the model recovers (bottom panels), assigned the same colour for each unique source object for Au14.

Once a set of stars have been assigned $P_{\rm a}$ values by the NNM, other physical parameters could be used to disentangle the contribution of false-positive identifications. Since the kinematics and binding energy have not been used in the NNMs, the only correlation with $P_{\rm a}$ will be due to innate correlations with the metallicity distribution and chemical abundances. As an example, the upper panels of Fig. \ref{fig:h14_source_recovery} displays stream-like stellar structures (yellow-green), and an inner core of accreted material (blue), which are recovered by the NNM, as displayed in the lower panels. In general, $\sim 60\%$ of the accreted stars in this galaxy are correctly labelled at $P_{\rm a} > 0.75$, as shown Table \ref{tab:networkperf}. 
The mis-classified in-situ stars displayed in Fig. \ref{fig:h14_source_recovery} (black points) are dominated by primarily old stars ($\tau_{\rm median} = 10.7 \rm ~Gyr$) that follow a circularity distribution centered around $\epsilon_{\rm J, median} = -0.08$. These stars are both chemically and kinematically much more similar to the accreted population than that of the global disc.
Other clustering of individual accreted substructures could be more evident with higher-parameter post-processing method such as streamfinding algorithms, or with more general clustering approaches, which can be applied after the NNM classifcation.

Au14 provides a good example of the potential for kinematic disentangling, and is discussed in more detail below.
Other galaxies may similarly provide excellent environments for distinction between true and false positive and negative assignments by the NNM. In-depth post-processing of NNM results that rely on kinematic differences between accreted and in-situ populations are beyond the scope of this work, which is primarily focused on the application of NNM techniques towards chemical information. Nevertheless, we analyse three examples of good and poor performance in order to assess the physical reasons behind these  behaviours. The other galaxies in the sample can be generally characterized as following the same pitfalls as these examples, regardless of galaxy-specific details - either performing well, over-estimation of accreted material, or over-estimation of in-situ material.

\subsection*{Au14}
\label{au14_sec}

Au14 is an example of a galaxy for which the NNMs perform well. We find that the star particles that NNMs assign $P_{\rm a} > 0.75$ in this galaxy's stellar disc are 80\% likely to be  true-positive accreted stars. The selected particles contain 59\% of all true accreted star particles.
The training data set we create for this galaxy is dominated by contributions from the stellar halo, and high-$M_{\rm star}$ satellites. Lower-$M_{\rm star}$ objects do not contribute enough stars to fill their allotment of the training set, as there are not enough satellite objects in this mass range at $z=0$. This leads to a small, but non-negligible fraction of false-negative classifications, though not enough to significantly impact the network performance measures.

\begin{figure*} 
    \includegraphics[width=2\columnwidth]{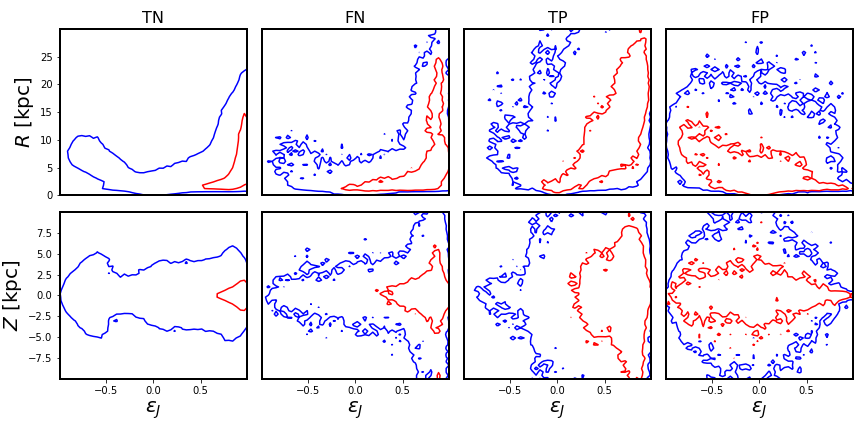}
    \caption{Projected radial distance, $R$, (top panels) on the rotation plane and disc height, $Z$, (bottom panels) as a function of circularity for the four classifications of star particles in our network evaluation of Au14: True-negative (TN), false-negative (FN), true-positive (TP) and false-positive(FP). Contours contain 50\% and 90\% of the total stellar mass of each classification.}
    \label{fig:Au14_epsj_rad}
\end{figure*}

\begin{figure*} 
   \includegraphics[width=2\columnwidth]{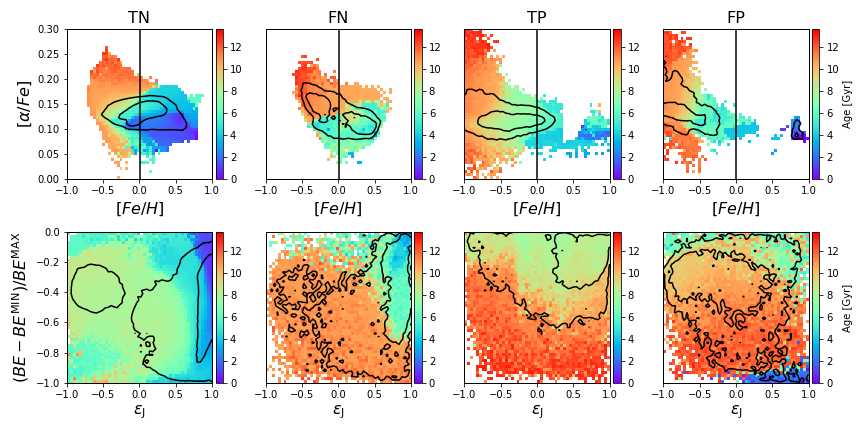}
   \caption{Upper panels: Histogram of $[\alpha/\rm Fe]$ vs $[\rm Fe/\rm H]$ for the True Negative (TN), False Negative (FN), True Positive (TP), and False Positive (FP) selections of star particles from Au14, coloured by the median particle age in each bin. The median galactic [Fe/H] is displayed with a vertical black line.
   Bottom panels: Histogram of binding energy (BE), rescaled by the highest and lowest values in the selection of star particles, and $\epsilon_{\rm J}$. Bins are coloured by the median age of the star particles they contain.
   Contours contain 50\% and 90\% of the total mass of star particles.}
   \label{fig:Au14_composite}
\end{figure*}

This galaxy also provides an excellent example for the use of kinematic data to augment the NNM results. Fig. \ref{fig:Au14_epsj_rad} displays the distributions of the kinematic parameter, $\epsilon_J$ and projected radial distance, $R$ and height, $Z$, in each network-assigned classification. The false-positive population  (fourth column) is distinct from the true-positive star particles (third column) primarily due to its lower circularity and central concentration, globally. This is due to the fact that, as discussed in \citet{G_mez_2017}, Au14 posses a very prominent ex-situ disc component. From this figure it is clear avoiding the central regions,  with $R < 5$ kpc, would improve the NNM's selection of accreted star particles. 
False-positive are distributed more evenly in $Z$ than accreted stars. 
These stars were misclassified due to their elevated [$\alpha$/Fe], and lower [Fe/H], as they tend to be older. This can be appreciated from  Fig.\ref{fig:Au14_composite} where [$\alpha$/Fe] are displayed together with the binding energy as a function of the angular momentum along z-axis, $L_{\rm z}$.

Those stars classified as false-negative have been misplaced  because they are more concentrated to the  central region with low $Z$. The upper panel of Fig. \ref{fig:Au14_composite} shows that the chemical abundances resembled better those of true in situ or true-negative stars. 
The lower panels of the figure demonstrate that although there is no kinematic information provided to the NNM, the subgroups formed by comparing the NNM-assigned label with the ground-truth populate phase space differently. Star particles labelled as in-situ (TN and FN) have more dominant young populations and are clearly supported by rotation, with only small counter-rotating elements. Star particles labelled as accreted (TP and FP) are predominantly older and less rotationally-supported, forming a thicker disc in the case of the true-positive selection, and contributing to the bulge in the false-positive case.

Other galaxies may not be as well-segregated, and would potentially require more complicated methods to separate correctly-labelled star particles from those that are incorrect. In this case, however, the boundaries are quite clear.

\subsection*{Au22}

\begin{figure*} 
    \includegraphics[width=2\columnwidth]{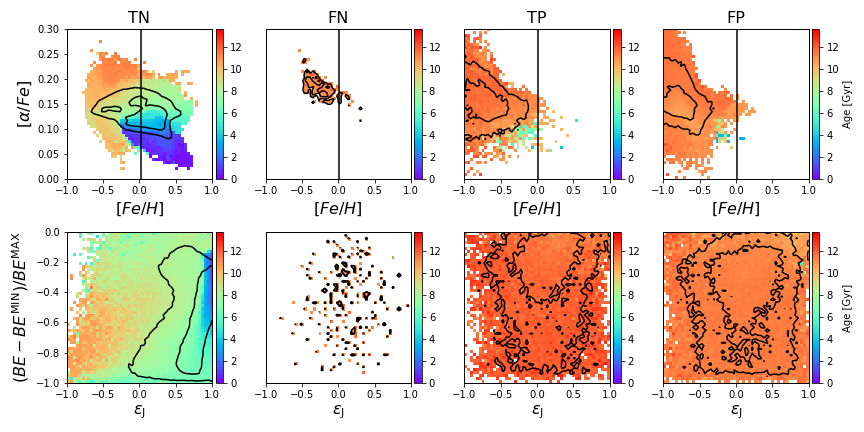}
    \caption{Upper panels: Histogram of $[\alpha/\rm Fe]$ vs $[\rm Fe/\rm H]$ for the True Negative (TN), False Negative (FN), True Positive (TP), and False Positive (FP) selections of star particles from Au22, coloured by the median particle age in each bin.  The median galactic [Fe/H] is displayed with a vertical black line.
    Lower Panels: Histogram of binding energy (BE), rescaled by the highest and lowest values in the selection of star particles, and $\epsilon_{\rm J}$. Bins are coloured by the median age of the star particles they contain.
   Contours contain 50\% and 90\% of the total mass of star particles.}
    \label{fig:Au22_composite}
\end{figure*}

Au22 is an example of a galaxy with a low $P(TP)$. While 97\% of the accreted stars are recovered at $P_{\rm a} > 0.75$, we would only be 26\% certain that a star the NNM has flagged as accreted has actually come from outside the primary galaxy. Its $f_{\rm acc}$ of 2\% falls at roughly half the mean false-positive rate, which can explain the particularly poor $P(TP)$ given such a high recovery fraction. 
The galaxy's trained NNM, however, also yields a large number of false-positive results. This is due to the presence of endo-debris, or star particles formed by gas stripped from orbiting satellite galaxies. This is demonstrated in the galaxy's complicated merger history. Between 13.5 to 11.6 Gyr ago, a satellite with nearly 9 times the gaseous mass of the central galaxy fell from 66 kpc to 13 kpc from the primary galaxy, losing nearly all of its gas in the process. Figure \ref{fig:Au22_composite} displays the distribution of $[\alpha/\rm Fe]$ vs $[\rm Fe/\rm H]$ and binding energy ($BE$) vs $L_{\rm z}$ for the various classification categories of stars in this galaxy. The false-positive $BE$-$L_{\rm z}$ distribution is similarly tightly bound in comparison with the true-positive population, and the age distribution of these mis-labelled stars falls off abruptly 7.5 Gyr ago, after peaking at roughly 11 Gyr, which is a distribution we would not expect from false-positives that are randomly mis-assigned from the in-situ population. This can be explained with endo-debris, or technically-in-situ stellar material formed from gas that was accreted from an infalling satellite, as defined in \cite{Tissera_2012}. If this were the true source of the false-positives, we would expect the age distribution of false-positive stars to be tied closely to the age distribution of true-positive accreted stars. From Fig. \ref{fig:Au22_composite}, we can see the similarities in both the metallicity and age distributions in the true- and false-positive selections. 

This example highlights the impact of gas-rich mergers on the performance of our method. If a galaxy has experienced such a merger, the endo-debris may be detected as accreted stellar material, despite ostensibly having formed in-situ, from gaseous debris mixed with the matter in the disc.

\subsection*{Au20}
\label{au20}

\begin{figure*} 
    \includegraphics[width=2\columnwidth]{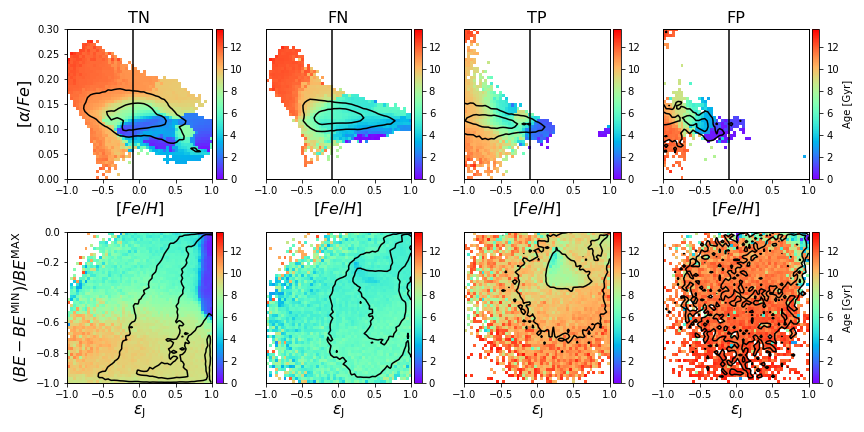}
    \caption{Upper panels: Histogram of $[\alpha/\rm Fe]$ vs $[\rm Fe/\rm H]$ for the True Negative (TN), False Negative (FN), True Positive (TP), and False Positive (FP) selections of star particles from Au20, coloured by the median particle age in each bin.  The median galactic [Fe/H] is displayed with a vertical black line.
    Lower Panels: Histogram of binding energy (BE), rescaled by the highest and lowest values in the selection of star particles, and $\epsilon_{\rm J}$. Bins are coloured by the median age of the star particles they contain.
   Contours contain 50\% and 90\% of the total mass of star particles.}
    \label{fig:Au20_composite}
\end{figure*}

Au20 is an example of a galaxy with a low $f_{\rm recov}$ in conjunction with a large reservoir of stars from massive satellites in its local environment, which comprise the majority of the training set candidate stars for this galaxy. Despite the presence of this reservoir of training stars, a massive majority of the misidentified false-negative stars originate from an object with a peak stellar mass (defined in \ref{network-training-subsection})  of $M_{\rm tot} = 10^{11.10} M_{\sun}$. This object has been orbiting the central galaxy for over 11 Gyr, and merged only 2.5 Gyr ago with a stellar mass of $M_{\star} = 10^{10.16} M_{\sun}$. Figure \ref{fig:Au20_composite} demonstrates that the distribution of $[\alpha/\rm Fe]$ vs $[\rm Fe/\rm H]$ values for the false-negative selection bears many more similarities to that of the true-negative sample, indicating that the accreted stars from this massive object are chemically similar to those in a central galaxy, not a satellite. 
As this example shows, if a merger takes place with a  galaxy with similar stellar mass, the accreted stars may be chemically indistinguishable from those that belong to our primary galaxy.

\begin{figure} 
    \includegraphics[width=\columnwidth]{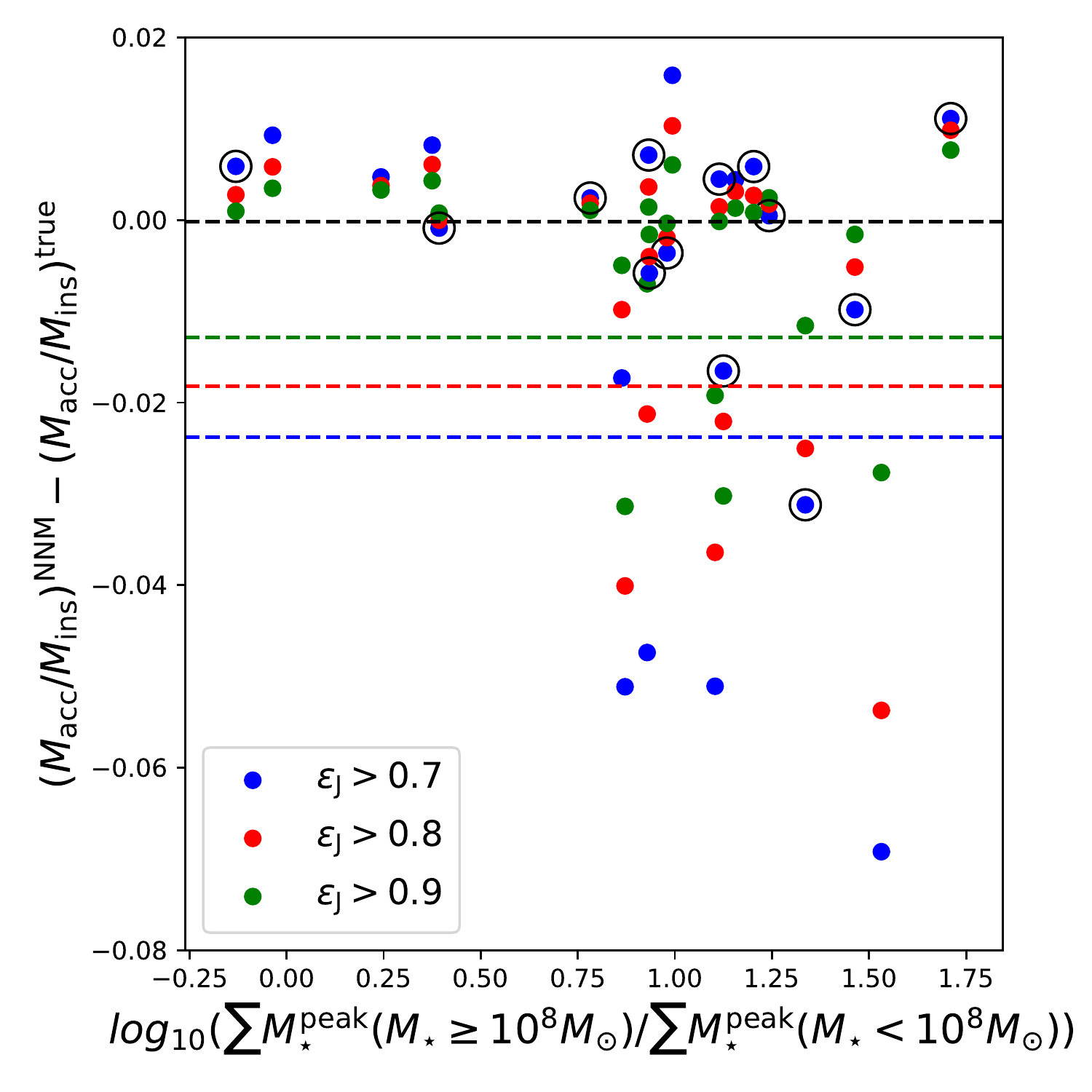}
    \caption{The difference between the true fraction of accreted to in-situ stellar mass and the fraction calculated with NNM results as a function of the ratio of the sum of the surviving satellite galaxies' peak stellar masses with $M_{\star} \geq 10^{8} M_{\odot}$ to that contained in galaxies less massive. The accreted and in-situ masses from the NNMs were taken from the output, requiring $P_{\rm a} > 0.75$ to be considered as accreted stars, and $P_{\rm a} < 0.5$ for in-situ. The mean differences at each circularity limit are displayed as coloured dashed lines. The median difference at $\epsilon_{\rm J} > 0.7$ is shown as a black dashed line. Galaxies that have been previously referred to as "good" are marked with a circle.}
    \label{fig:dMxMi_ej}
\end{figure}

In fact  we have found that \maga~  generally recover  accreted stellar populations which are mainly rotationally dominated but also extended to lower $\epsilon_{\rm J}$ in agreement with \cite{G_mez_2017} (Fig.5).
We further investigate this by estimating the fraction of accreted stars over in-situ at $\epsilon_{\rm J}$ selections above 0.7, 0.8, and 0.9, that are recovered by \maga~. This is shown in Fig. \ref{fig:dMxMi_ej}. The NNM results were calculated assuming every star particle with $P_{\rm a} > 0.75$ is accreted, and every particle with $P_{\rm a} < 0.5$ was formed in-situ. On average, \maga results underestimate the fraction of accreted to in-situ stars by up to two per cent. The black dashed line, corresponding to the median difference in $M_{\rm acc}/M_{\rm ins}$ values at $\epsilon_{\rm J} > 0.7$ across our galaxy selection, shows no significant shift. This implies the deviation is driven by several outliers, as opposed to being a systematic underestimation of the fraction of accreted star particles. Selections based on circularity, or other non-chemical parameters have the potential to provide crucial post-processing in low P(TP) situations.


\subsection{Impact of Galactic Environment}


An interesting aspect of our NNM is that, once trained, it does not require any external data for further training. This means that it could be readily applied  to observational data without the need of a companion simulation suite. Note that the models used for training are MW analogs that sample a wide range of different possible assembly histories. We have also shown that the  method is very successful for half (12) of the Auriga simulated galaxies that we have evaluated, providing us with $f_{\rm recov} > 0.5$ and $P(TP) > 0.5$ for $P_{\rm a} > 0.75$. In these galaxies, the training selection of stars encompasses enough of the variation present in the galactic disc for in-situ and accreted material to be distinguished and separated. However, for the other half of our sample, the results yielded either low  $f_{\rm recov}$ and/or $P(TP)$ (below 50\%).

The under perfomance of our NNM in these cases could be associated with one of the principal components of our method: {\it the local galactic environment} from which the the training set is obtained. To explore this, we  quantified the local environment by computing the probability distribution function (PDF) of the stellar mass of nearby satellites and contributors to the outer stellar halo (i.e. stars beyond 25 kpc as described in Section 3.3). While an evaluation suite of Auriga galaxies has allowed us to determine that the method will distinguish between accreted and in-situ stars in general, we need to compare the surviving satellite mass distributions to draw any further conclusions about performance with the MW itself.

\begin{figure} 
    \includegraphics[width=0.99\columnwidth]{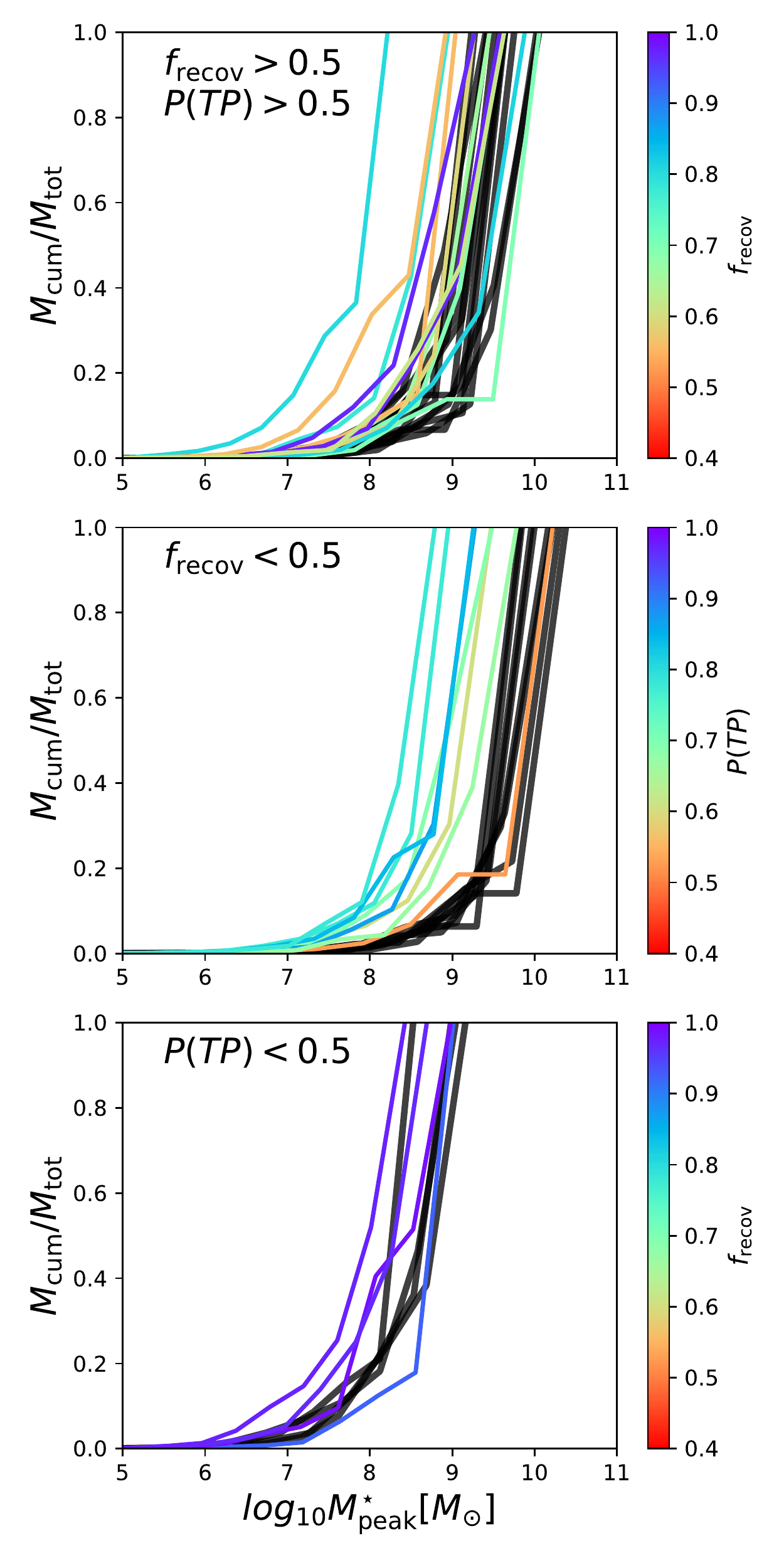}
    \caption{
    Cumulative mass fractions of the peak masses of satellites, as well as the peak masses of objects that have contributed to the outer stellar halo (identified by a unique peak mass IDs) as a function of the stellar mass of the source, $M^{\rm peak}_{\star}$, for each Auriga galaxy, separated by NNM performance. The distributions are shown for stars in the environments (coloured by either $f_{\rm recov}$ or $P(TP)$) and  for  the accreted star particles in the corresponding galactic discs (black lines).
    Galaxies for which an NNM achieves $P(TP) > 0.5 | P_{\rm a} > 0.75$ are in the top panel, galaxies for which an NNM provides $f_{\rm rec} < 0.5 | P_{\rm a} > 0.75$ are located in the middle panel, and  galaxies for which an NNM provides $P(TP) < 0.5 | P_{\rm a} > 0.75$ are shown in the bottom panel.}
    \label{fig:sat_source_comp}
\end{figure}

Figure \ref{fig:sat_source_comp} displays the cumulative mass fractions of each surviving satellite's peak mass, plus the peak masses of objects that have contributed star particles to the outer stellar halo (identified by unique peak mass IDs), as a function of the stellar mass of their source satellite, $M^{\rm peak}_{\star}$, for each Auriga galaxy, separated by NNM performance.
The cumulative distributions are shown as lines coloured by either $f_{\rm recov}$ or $P(TP)$. 
The black lines represent the equivalent  distributions for accreted stars in the discs for the same NNM performance selections.
Galaxies for which a NNM achieves $P(TP) > 0.5 | P_{\rm a} > 0.75$ are in the top panel, galaxies for which an NNM provides $f_{\rm rec} < 0.5 | P_{\rm a} > 0.75$ are located in the middle panel, and  galaxies for which an NNM provides $P(TP) < 0.5 | P_{\rm a} > 0.75$ are located in the bottom panel.

Auriga systems for which  both $f_{\rm recov}$ and $P(TP)$ are above 0.5 (top panel: Au5, Au6, Au12, Au13, Au14, Au15, Au16, Au23, Au24, Au26) tend to have accreted stellar particles in the disc region originated in sources with similar $M^{\rm peak}_{\star}$  distribution to their environment (from which the training set are selected). Galaxies for which NNMs result in a lower recovery fraction (middle panel, Au2, Au3, Au4, Au7, Au19, Au20, Au21, Au27, Au28) tend to have accreted stellar particles in the disc region that originated from  higher $M^{\rm peak}_{\star}$ compared to those that contribute to training sets. 
 We would expect this given the difficulty NNMs experience distinguishing between stars accreted from extremely massive objects and those formed in-situ in a massive galaxy, as was discussed previously for Au20. These galaxies typically host significant ex-situ discs \citep{G_mez_2017}. We find that only one of the galaxies in this performance selection follows a similar distribution of source masses above $10^{8} M_{\sun}$, while the rest have training set populations that fail to represent the highest mass contributors to their respective stellar discs. 
This can be understood from the results reported by \citet{Monachesi_2019} where the assembly histories of the  stellar haloes of the Auriga galaxies are studied in detail. Our findings for this particular subset of galaxies, with low $f_{\rm recov}$, which have  environments formed mainly from small satellites, agree with their results. However, more massive satellites have contributed to their discs, and hence to the evaluation set, with significant fractions of accreted stars. From Table 1 we can estimate $f_{\rm acc} =[0.11, 0.36]$ and a mean value of $f_{\rm acc} = 0.22$. Our training set missed their contributions because these stars tend to be concentrated around the discs within $\sim 20$~kpc as shown in figure 4 of Monachesi et al. Hence, they are missed in our training set by construction. 

 Finally,  galaxies for which NNMs result in lower P(TP) values (bottom panel, Au9, Au10, Au17, Au18, Au22) have distributions that match very well, but tend to lack star particles from higher-mass sources in both the population of accreted disc star particles, and the star particles in the halo and surviving satellites \citep[see also figure 8 in ][]{Monachesi_2019}. Due to the lack of massive contributors to the discs in these galaxies, the fraction of accreted star particles is systematically lower. In fact, the mean values is $f_{\rm acc} = 0.03$.  As the false-positive rate is comparable to the fraction of accreted star particles to find in the disc, the P(TP) will be below 50\%. This false-positive rate could also be driven partially by endo-debris, as described in Section \ref{au20}.


\subsection{Accretion Source Recovery: The GES analogue test}

To  prove the predictive power of our algorithm, we focus on the well-identified massive merger that the MW had roughly 10 billion years ago,  the Gaia-Enceladus (GES) galaxy  \citep{Helmi_2018, Belokurov_2018}. The stellar debris from this event covers nearly the full sky and, at the time of the merger, the GES is estimated to have contained nearly a quarter of the mass of the progenitor of the MW. One of the goals for our NNM training and evaluation method is to be able to detect debris from disrupted satellites, such as the GES, in the MW disc.

To this end, we generated a "GES-mass proxy" sample for each tested Auriga halo. This proxy is created by sampling old stars ($\tau > 10$ Gyr) from the satellite galaxy with the highest $M_{\rm star}$ within the virial radius of the central galaxy \citep{Helmi_2018}. This definition, by necessity, neglects most of the physical and kinematic characteristics of the GES itself. A true analogue to the GES has been  demonstrated by \cite{Bignone_2019} to be rare and none of the Auriga initial conditions have been designed to include a GES event. By instead sampling appropriately-aged stars from massive surviving satellites, we allow ourselves to obtain results from every Auriga volume. This approach will approximate an object composed of stars formed before the GES would have ceased star formation, though we will neglect the effects that accreted gas may have had on the development of endo-debris. The NNMs for these galaxies are trained with the same data set as previously, with the exception that  the stars associated with the selected GES-proxy have been removed, to prevent foreknowledge about any of the stars in the structure. Since any accreted stars in GES-like structures will be old stars from a single satellite source, our NNM's performance in this test should correspond with future performance in the discovery of large accreted components in the MW disc. While the isolation of stars belonging to GES-like objects or substructures will require additional post-processing, in an application of \maga~ with real data, this test will be indicative of ideal performance.  While these results are based on the mass-metallicity relation of the Auriga simulations, as long as {\it a} clear mass-metallicity relation exists in the training and evaluation populations the NNM should be able to identify accreted stars.

Table \ref{tab:enc_test} displays our NNM's performance with a GES-proxy from the halo of each Auriga galaxy. The stellar mass ($M_{\rm star}$) of the GES-proxy, and the recovery of the stars at three $P_{\rm a}$ thresholds are displayed. In a majority ($18/24$) of our galaxies, more than 80\% of the star particles that make up the GES proxy are recovered, for $P_{\rm a} > 0.75$. With a lower threshold of $P_{\rm a} > 0.50$, every GES-proxy is recovered above 50\%.

\begin{table}
    \caption{$f_{\rm recov}$ for a GES proxy constituted of old stars ($\tau > 10 ~\rm Gyr$) from a single satellite source located in  each Auriga galactic halo. The Auriga Id and stellar mass of the GES-proxy object is included ( first and second columns). The $f_{\rm recov}$ for  different thresholds $t$ are given (3-5 columns).}
    \resizebox{0.5\textwidth}{!}{%
    \begin{tabular}{ccccc}
    \hline
        Halo & $ \rm log_{10} M_{\star}/M_{\odot}$ & $f_{\rm recov}|P_{\rm a} > 0.90$ & $f_{\rm recov}|P_{\rm a} > 0.75$ & $f_{\rm recov}|P_{\rm a} > 0.50$ \\
        \hline 
        2 & 9.35 & 0.15 & 0.45 & 0.63 \\
        3 & 8.64 & 0.67 & 0.89 & 0.97 \\
        4 & 8.75 & 0.34 & 0.95 & 0.99 \\
        5 & 8.17 & 0.16 & 0.73 & 0.95 \\
        6 & 7.98 & 0.06 & 0.40 & 0.89 \\
        7 & 8.07 & 0.04 & 0.13 & 0.52 \\
        9 & 8.30 & 0.46 & 0.82 & 0.93 \\
        10 & 8.23 & 0.32 & 0.91 & 1.00 \\
        12 & 8.45 & 0.26 & 0.83 & 0.93 \\
        13 & 8.07 & 0.34 & 0.88 & 0.98 \\
        14 & 7.94 & 0.22 & 0.90 & 0.97 \\
        15 & 9.09 & 0.09 & 0.50 & 0.84 \\
        16 & 8.46 & 0.28 & 0.99 & 1.00 \\
        17 & 7.90 & 0.85 & 0.99 & 1.00 \\
        18 & 8.52 & 0.24 & 0.48 & 0.68 \\
        19 & 8.23 & 0.16 & 0.83 & 0.99 \\
        20 & 8.27 & 0.22 & 0.72 & 0.98 \\
        21 & 8.69 & 0.04 & 0.64 & 0.95 \\
        22 & 7.78 & 0.72 & 0.78 & 0.85 \\
        23 & 8.64 & 0.07 & 0.24 & 0.81 \\
        24 & 8.67 & 0.08 & 0.45 & 0.81 \\
        26 & 8.76 & 0.51 & 0.83 & 0.92 \\
        27 & 8.43 & 0.50 & 0.93 & 0.98 \\
        28 & 8.52 & 0.92 & 0.98 & 1.00 \\
        \hline
    \end{tabular}
    }
\label{tab:enc_test}
\end{table}


\section{Cross-NNM Performance Tests}
\label{sec:cross_perf}

While our method of NNM training leads to models that are specialized towards each galaxy's specific environment, certain environments provide a well-rounded training sample of potential accreted stars that is applicable more broadly. Table \ref{tab:cross_ptp} and Table \ref{tab:cross_frec} display the P(TP) and $f_{\rm rec}$ of each NNM trained in a given Auriga environment when applied to each Auriga galactic disc. Cells are coloured based on their values. Blue cells indicate values above 0.75, green indicates values above 0.5, and pink indicates values below 0.5. Columns that consist of primarily green and blue in both tables demonstrates excellent NNM performance across the set of galactic discs.

Table \ref{tab:cross_frec} presents two clear columns representing galaxies which provided poor NNM training sets. Au7 and Au20 both share recent mergers.
In both of these cases, these events create a disc environment that is composed of $>30\%$ accreted star particles. In the case of Au7, a majority of the false-negative star particles originate from one of two merger events from 3.97 Gyr and 1.34 Gyr ago. Au20 presents a very similar situation. A single merger event that took place 2.48 Gyr ago contributed a majority of the false-negative disc stars. In both these cases, if we select only true in-situ star particles from the disc to create the in-situ half of the training set, the false-negatives disappear. This implies that the mislabelling is not due to chemical similarity between these accreted stars and the in-situ population of the galactic disc. This is a demonstration of the weakness of our observationally-motivated training set selection scheme, which is predicated on the disc being principally composed of in-situ star particles, and was designed around an imposed inability to remove accreted contamination from the disc a priori. This is evidenced by the improved performance other NNMs display when executed on both Au7 and Au20.

\begin{table*}
    \caption{Cross-performance table for neural network $P(TP)|P_{\rm a}>0.75$ for NNMs trained on selected Auriga galaxies (columns) evaluating the star particles from the same set of galaxies (rows). Cells are coloured based on performance thresholds, with pink indicating a P(TP) < 0.5, green a P(TP) < 0.75, and blue a P(TP) > 0.75.}
    \resizebox{\textwidth}{!}{%
    \begin{tabular}{lllllllllllllllllllllllllll}
    \hline
        Validation\textbackslash{}Training & 2 & 3 & 4 & 5 & 6 & 7 & 9 & 10 & 12 & 13 & 14 & 15 & 16 & 17 & 18 & 19 & 20 & 21 & 22 & 23 & 24 & 26 & 27 & 28 \\
        \hline
        2 & \cellcolor{magenta}0.4 & \cellcolor{green}0.6 & \cellcolor{green}0.5 & \cellcolor{green}0.6 & \cellcolor{cyan}0.8 & \cellcolor{green}0.7 & \cellcolor{green}0.6 & \cellcolor{green}0.7 & \cellcolor{green}0.6 & \cellcolor{green}0.7 & \cellcolor{magenta}0.5 & \cellcolor{green}0.6 & \cellcolor{green}0.6 & \cellcolor{magenta}0.5 & \cellcolor{green}0.6 & \cellcolor{green}0.6 & \cellcolor{green}0.7 & \cellcolor{green}0.5 & \cellcolor{green}0.6 & \cellcolor{green}0.7 & \cellcolor{green}0.6 & \cellcolor{green}0.6 & \cellcolor{green}0.5 & \cellcolor{green}0.5 \\
        3 & \cellcolor{magenta}0.4 & \cellcolor{green}0.6 & \cellcolor{green}0.6 & \cellcolor{green}0.6 & \cellcolor{green}0.7 & \cellcolor{green}0.6 & \cellcolor{green}0.6 & \cellcolor{green}0.7 & \cellcolor{green}0.6 & \cellcolor{green}0.6 & \cellcolor{green}0.5 & \cellcolor{green}0.6 & \cellcolor{green}0.6 & \cellcolor{green}0.5 & \cellcolor{green}0.6 & \cellcolor{green}0.6 & \cellcolor{cyan}0.8 & \cellcolor{green}0.6 & \cellcolor{green}0.6 & \cellcolor{green}0.7 & \cellcolor{green}0.6 & \cellcolor{green}0.5 & \cellcolor{green}0.6 & \cellcolor{green}0.5 \\
        4 & \cellcolor{magenta}0.4 & \cellcolor{magenta}0.5 & \cellcolor{green}0.7 & \cellcolor{green}0.6 & \cellcolor{green}0.5 & \cellcolor{green}0.5 & \cellcolor{green}0.6 & \cellcolor{green}0.5 & \cellcolor{green}0.5 & \cellcolor{green}0.6 & \cellcolor{green}0.5 & \cellcolor{green}0.7 & \cellcolor{green}0.7 & \cellcolor{magenta}0.5 & \cellcolor{green}0.6 & \cellcolor{green}0.6 & \cellcolor{green}0.6 & \cellcolor{magenta}0.5 & \cellcolor{green}0.5 & \cellcolor{magenta}0.5 & \cellcolor{green}0.6 & \cellcolor{magenta}0.5 & \cellcolor{magenta}0.4 & \cellcolor{green}0.6 \\
        5 & \cellcolor{magenta}0.4 & \cellcolor{green}0.6 & \cellcolor{magenta}0.5 & \cellcolor{green}0.6 & \cellcolor{green}0.7 & \cellcolor{green}0.7 & \cellcolor{green}0.5 & \cellcolor{green}0.7 & \cellcolor{green}0.6 & \cellcolor{green}0.7 & \cellcolor{magenta}0.5 & \cellcolor{green}0.5 & \cellcolor{green}0.5 & \cellcolor{magenta}0.4 & \cellcolor{green}0.6 & \cellcolor{green}0.5 & \cellcolor{green}0.7 & \cellcolor{green}0.5 & \cellcolor{green}0.6 & \cellcolor{green}0.7 & \cellcolor{green}0.5 & \cellcolor{green}0.6 & \cellcolor{green}0.5 & \cellcolor{magenta}0.5 \\
        6 & \cellcolor{green}0.5 & \cellcolor{green}0.6 & \cellcolor{green}0.5 & \cellcolor{green}0.5 & \cellcolor{green}0.7 & \cellcolor{green}0.7 & \cellcolor{green}0.5 & \cellcolor{green}0.7 & \cellcolor{green}0.6 & \cellcolor{green}0.7 & \cellcolor{magenta}0.4 & \cellcolor{green}0.6 & \cellcolor{green}0.6 & \cellcolor{magenta}0.5 & \cellcolor{green}0.6 & \cellcolor{green}0.6 & \cellcolor{cyan}0.8 & \cellcolor{green}0.5 & \cellcolor{green}0.6 & \cellcolor{green}0.7 & \cellcolor{magenta}0.5 & \cellcolor{green}0.5 & \cellcolor{green}0.5 & \cellcolor{magenta}0.4 \\
        7 & \cellcolor{green}0.6 & \cellcolor{green}0.6 & \cellcolor{cyan}0.8 & \cellcolor{cyan}0.8 & \cellcolor{cyan}0.8 & \cellcolor{cyan}0.9 & \cellcolor{cyan}0.8 & \cellcolor{cyan}0.8 & \cellcolor{cyan}0.8 & \cellcolor{cyan}0.8 & \cellcolor{green}0.7 & \cellcolor{cyan}0.8 & \cellcolor{cyan}0.8 & \cellcolor{green}0.7 & \cellcolor{cyan}0.8 & \cellcolor{cyan}0.8 & \cellcolor{cyan}0.9 & \cellcolor{green}0.6 & \cellcolor{green}0.7 & \cellcolor{cyan}0.8 & \cellcolor{cyan}0.8 & \cellcolor{green}0.7 & \cellcolor{green}0.5 & \cellcolor{cyan}0.8 \\
        9 & \cellcolor{magenta}0.5 & \cellcolor{magenta}0.5 & \cellcolor{magenta}0.4 & \cellcolor{magenta}0.4 & \cellcolor{green}0.6 & \cellcolor{cyan}0.8 & \cellcolor{magenta}0.4 & \cellcolor{green}0.6 & \cellcolor{green}0.5 & \cellcolor{green}0.7 & \cellcolor{magenta}0.4 & \cellcolor{green}0.6 & \cellcolor{magenta}0.5 & \cellcolor{magenta}0.4 & \cellcolor{magenta}0.5 & \cellcolor{magenta}0.5 & \cellcolor{green}0.7 & \cellcolor{magenta}0.4 & \cellcolor{green}0.5 & \cellcolor{green}0.6 & \cellcolor{magenta}0.4 & \cellcolor{magenta}0.4 & \cellcolor{magenta}0.4 & \cellcolor{magenta}0.4 \\
        10 & \cellcolor{magenta}0.2 & \cellcolor{magenta}0.3 & \cellcolor{magenta}0.2 & \cellcolor{magenta}0.2 & \cellcolor{magenta}0.4 & \cellcolor{green}0.6 & \cellcolor{magenta}0.2 & \cellcolor{magenta}0.5 & \cellcolor{magenta}0.3 & \cellcolor{green}0.6 & \cellcolor{magenta}0.2 & \cellcolor{magenta}0.3 & \cellcolor{magenta}0.3 & \cellcolor{magenta}0.2 & \cellcolor{magenta}0.3 & \cellcolor{magenta}0.3 & \cellcolor{green}0.7 & \cellcolor{magenta}0.3 & \cellcolor{magenta}0.3 & \cellcolor{magenta}0.4 & \cellcolor{magenta}0.2 & \cellcolor{magenta}0.3 & \cellcolor{magenta}0.2 & \cellcolor{magenta}0.2 \\
        12 & \cellcolor{magenta}0.4 & \cellcolor{green}0.6 & \cellcolor{green}0.6 & \cellcolor{green}0.6 & \cellcolor{green}0.7 & \cellcolor{green}0.7 & \cellcolor{green}0.6 & \cellcolor{green}0.7 & \cellcolor{green}0.6 & \cellcolor{green}0.7 & \cellcolor{magenta}0.5 & \cellcolor{green}0.6 & \cellcolor{green}0.6 & \cellcolor{magenta}0.5 & \cellcolor{green}0.6 & \cellcolor{green}0.6 & \cellcolor{green}0.7 & \cellcolor{green}0.6 & \cellcolor{green}0.6 & \cellcolor{green}0.7 & \cellcolor{green}0.6 & \cellcolor{magenta}0.5 & \cellcolor{green}0.6 & \cellcolor{green}0.6 \\
        13 & \cellcolor{magenta}0.4 & \cellcolor{green}0.6 & \cellcolor{magenta}0.5 & \cellcolor{green}0.5 & \cellcolor{green}0.7 & \cellcolor{green}0.7 & \cellcolor{green}0.5 & \cellcolor{green}0.7 & \cellcolor{green}0.6 & \cellcolor{green}0.7 & \cellcolor{magenta}0.4 & \cellcolor{green}0.5 & \cellcolor{green}0.6 & \cellcolor{magenta}0.4 & \cellcolor{green}0.6 & \cellcolor{magenta}0.5 & \cellcolor{green}0.7 & \cellcolor{green}0.5 & \cellcolor{green}0.5 & \cellcolor{green}0.6 & \cellcolor{green}0.5 & \cellcolor{magenta}0.5 & \cellcolor{green}0.5 & \cellcolor{magenta}0.5 \\
        14 & \cellcolor{green}0.6 & \cellcolor{cyan}0.8 & \cellcolor{green}0.6 & \cellcolor{cyan}0.8 & \cellcolor{cyan}0.9 & \cellcolor{cyan}0.9 & \cellcolor{green}0.7 & \cellcolor{cyan}0.9 & \cellcolor{cyan}0.8 & \cellcolor{cyan}0.9 & \cellcolor{green}0.7 & \cellcolor{green}0.7 & \cellcolor{cyan}0.8 & \cellcolor{green}0.7 & \cellcolor{cyan}0.8 & \cellcolor{green}0.7 & \cellcolor{cyan}0.9 & \cellcolor{green}0.7 & \cellcolor{cyan}0.8 & \cellcolor{cyan}0.8 & \cellcolor{green}0.7 & \cellcolor{cyan}0.8 & \cellcolor{green}0.7 & \cellcolor{green}0.7 \\
        15 & \cellcolor{magenta}0.2 & \cellcolor{magenta}0.3 & \cellcolor{green}0.6 & \cellcolor{magenta}0.5 & \cellcolor{magenta}0.4 & \cellcolor{green}0.5 & \cellcolor{magenta}0.4 & \cellcolor{magenta}0.4 & \cellcolor{magenta}0.4 & \cellcolor{green}0.5 & \cellcolor{magenta}0.3 & \cellcolor{green}0.6 & \cellcolor{magenta}0.5 & \cellcolor{magenta}0.3 & \cellcolor{magenta}0.5 & \cellcolor{green}0.5 & \cellcolor{green}0.6 & \cellcolor{magenta}0.3 & \cellcolor{magenta}0.3 & \cellcolor{magenta}0.3 & \cellcolor{magenta}0.5 & \cellcolor{magenta}0.2 & \cellcolor{magenta}0.3 & \cellcolor{magenta}0.4 \\
        16 & \cellcolor{magenta}0.4 & \cellcolor{magenta}0.5 & \cellcolor{magenta}0.5 & \cellcolor{magenta}0.5 & \cellcolor{green}0.7 & \cellcolor{green}0.6 & \cellcolor{magenta}0.5 & \cellcolor{green}0.7 & \cellcolor{green}0.5 & \cellcolor{green}0.6 & \cellcolor{magenta}0.4 & \cellcolor{green}0.5 & \cellcolor{green}0.5 & \cellcolor{magenta}0.3 & \cellcolor{green}0.6 & \cellcolor{magenta}0.5 & \cellcolor{green}0.7 & \cellcolor{magenta}0.4 & \cellcolor{magenta}0.5 & \cellcolor{green}0.6 & \cellcolor{magenta}0.5 & \cellcolor{magenta}0.5 & \cellcolor{magenta}0.4 & \cellcolor{magenta}0.4 \\
        17 & \cellcolor{magenta}0.3 & \cellcolor{magenta}0.3 & \cellcolor{magenta}0.2 & \cellcolor{magenta}0.3 & \cellcolor{magenta}0.4 & \cellcolor{green}0.7 & \cellcolor{magenta}0.2 & \cellcolor{magenta}0.5 & \cellcolor{magenta}0.4 & \cellcolor{green}0.7 & \cellcolor{magenta}0.3 & \cellcolor{magenta}0.3 & \cellcolor{magenta}0.3 & \cellcolor{magenta}0.3 & \cellcolor{magenta}0.3 & \cellcolor{magenta}0.3 & \cellcolor{green}0.7 & \cellcolor{magenta}0.3 & \cellcolor{magenta}0.3 & \cellcolor{magenta}0.4 & \cellcolor{magenta}0.2 & \cellcolor{magenta}0.2 & \cellcolor{magenta}0.2 & \cellcolor{magenta}0.2 \\
        18 & \cellcolor{magenta}0.3 & \cellcolor{magenta}0.4 & \cellcolor{magenta}0.3 & \cellcolor{magenta}0.4 & \cellcolor{green}0.5 & \cellcolor{green}0.6 & \cellcolor{magenta}0.3 & \cellcolor{green}0.6 & \cellcolor{magenta}0.4 & \cellcolor{green}0.6 & \cellcolor{magenta}0.3 & \cellcolor{magenta}0.4 & \cellcolor{magenta}0.4 & \cellcolor{magenta}0.3 & \cellcolor{magenta}0.4 & \cellcolor{magenta}0.4 & \cellcolor{green}0.7 & \cellcolor{magenta}0.4 & \cellcolor{magenta}0.4 & \cellcolor{magenta}0.5 & \cellcolor{magenta}0.3 & \cellcolor{magenta}0.3 & \cellcolor{magenta}0.3 & \cellcolor{magenta}0.3 \\
        19 & \cellcolor{magenta}0.5 & \cellcolor{green}0.6 & \cellcolor{green}0.6 & \cellcolor{green}0.7 & \cellcolor{cyan}0.8 & \cellcolor{cyan}0.8 & \cellcolor{green}0.7 & \cellcolor{cyan}0.8 & \cellcolor{green}0.7 & \cellcolor{cyan}0.8 & \cellcolor{green}0.6 & \cellcolor{green}0.7 & \cellcolor{green}0.7 & \cellcolor{green}0.5 & \cellcolor{green}0.7 & \cellcolor{green}0.7 & \cellcolor{cyan}0.8 & \cellcolor{green}0.6 & \cellcolor{green}0.7 & \cellcolor{cyan}0.8 & \cellcolor{green}0.6 & \cellcolor{green}0.6 & \cellcolor{green}0.6 & \cellcolor{green}0.6 \\
        20 & \cellcolor{green}0.7 & \cellcolor{green}0.7 & \cellcolor{green}0.6 & \cellcolor{green}0.7 & \cellcolor{cyan}0.9 & \cellcolor{cyan}0.8 & \cellcolor{green}0.7 & \cellcolor{cyan}0.8 & \cellcolor{cyan}0.8 & \cellcolor{cyan}0.8 & \cellcolor{cyan}0.8 & \cellcolor{green}0.6 & \cellcolor{green}0.7 & \cellcolor{green}0.7 & \cellcolor{green}0.7 & \cellcolor{green}0.7 & \cellcolor{cyan}0.9 & \cellcolor{green}0.7 & \cellcolor{cyan}0.8 & \cellcolor{cyan}0.9 & \cellcolor{green}0.7 & \cellcolor{cyan}0.8 & \cellcolor{green}0.7 & \cellcolor{green}0.7 \\
        21 & \cellcolor{green}0.5 & \cellcolor{green}0.7 & \cellcolor{green}0.6 & \cellcolor{green}0.6 & \cellcolor{cyan}0.8 & \cellcolor{green}0.7 & \cellcolor{green}0.6 & \cellcolor{cyan}0.8 & \cellcolor{green}0.6 & \cellcolor{green}0.7 & \cellcolor{green}0.5 & \cellcolor{green}0.7 & \cellcolor{green}0.7 & \cellcolor{green}0.5 & \cellcolor{green}0.7 & \cellcolor{green}0.6 & \cellcolor{cyan}0.8 & \cellcolor{green}0.6 & \cellcolor{green}0.6 & \cellcolor{green}0.7 & \cellcolor{green}0.6 & \cellcolor{green}0.6 & \cellcolor{green}0.6 & \cellcolor{green}0.6 \\
        22 & \cellcolor{magenta}0.3 & \cellcolor{magenta}0.3 & \cellcolor{magenta}0.2 & \cellcolor{magenta}0.3 & \cellcolor{magenta}0.4 & \cellcolor{green}0.5 & \cellcolor{magenta}0.2 & \cellcolor{magenta}0.4 & \cellcolor{magenta}0.3 & \cellcolor{magenta}0.4 & \cellcolor{magenta}0.2 & \cellcolor{magenta}0.3 & \cellcolor{magenta}0.3 & \cellcolor{magenta}0.3 & \cellcolor{magenta}0.3 & \cellcolor{magenta}0.3 & \cellcolor{green}0.5 & \cellcolor{magenta}0.3 & \cellcolor{magenta}0.3 & \cellcolor{magenta}0.4 & \cellcolor{magenta}0.2 & \cellcolor{magenta}0.3 & \cellcolor{magenta}0.3 & \cellcolor{magenta}0.2 \\
        23 & \cellcolor{green}0.6 & \cellcolor{green}0.7 & \cellcolor{green}0.6 & \cellcolor{green}0.7 & \cellcolor{cyan}0.8 & \cellcolor{cyan}0.9 & \cellcolor{green}0.6 & \cellcolor{cyan}0.8 & \cellcolor{green}0.7 & \cellcolor{cyan}0.9 & \cellcolor{green}0.6 & \cellcolor{green}0.7 & \cellcolor{green}0.7 & \cellcolor{green}0.6 & \cellcolor{green}0.7 & \cellcolor{green}0.7 & \cellcolor{cyan}0.9 & \cellcolor{green}0.7 & \cellcolor{green}0.7 & \cellcolor{cyan}0.8 & \cellcolor{green}0.6 & \cellcolor{green}0.7 & \cellcolor{green}0.6 & \cellcolor{green}0.6 \\
        24 & \cellcolor{magenta}0.4 & \cellcolor{green}0.5 & \cellcolor{magenta}0.5 & \cellcolor{green}0.6 & \cellcolor{green}0.7 & \cellcolor{green}0.5 & \cellcolor{green}0.5 & \cellcolor{green}0.7 & \cellcolor{green}0.5 & \cellcolor{green}0.6 & \cellcolor{magenta}0.5 & \cellcolor{green}0.5 & \cellcolor{green}0.6 & \cellcolor{magenta}0.4 & \cellcolor{green}0.6 & \cellcolor{green}0.5 & \cellcolor{green}0.7 & \cellcolor{magenta}0.5 & \cellcolor{green}0.6 & \cellcolor{green}0.7 & \cellcolor{green}0.5 & \cellcolor{green}0.6 & \cellcolor{magenta}0.5 & \cellcolor{green}0.5 \\
        26 & \cellcolor{green}0.6 & \cellcolor{green}0.7 & \cellcolor{green}0.7 & \cellcolor{green}0.7 & \cellcolor{cyan}0.8 & \cellcolor{cyan}0.8 & \cellcolor{green}0.7 & \cellcolor{cyan}0.8 & \cellcolor{green}0.7 & \cellcolor{cyan}0.8 & \cellcolor{green}0.6 & \cellcolor{cyan}0.8 & \cellcolor{green}0.7 & \cellcolor{green}0.7 & \cellcolor{green}0.7 & \cellcolor{green}0.7 & \cellcolor{cyan}0.8 & \cellcolor{green}0.7 & \cellcolor{green}0.7 & \cellcolor{cyan}0.8 & \cellcolor{green}0.7 & \cellcolor{green}0.7 & \cellcolor{green}0.7 & \cellcolor{green}0.6 \\
        27 & \cellcolor{magenta}0.4 & \cellcolor{green}0.7 & \cellcolor{green}0.6 & \cellcolor{green}0.6 & \cellcolor{green}0.7 & \cellcolor{cyan}0.8 & \cellcolor{green}0.6 & \cellcolor{green}0.7 & \cellcolor{green}0.6 & \cellcolor{green}0.7 & \cellcolor{green}0.6 & \cellcolor{green}0.7 & \cellcolor{green}0.7 & \cellcolor{green}0.6 & \cellcolor{green}0.7 & \cellcolor{green}0.7 & \cellcolor{cyan}0.8 & \cellcolor{green}0.6 & \cellcolor{green}0.7 & \cellcolor{green}0.7 & \cellcolor{green}0.6 & \cellcolor{green}0.6 & \cellcolor{green}0.6 & \cellcolor{green}0.6 \\
        28 & \cellcolor{green}0.6 & \cellcolor{green}0.7 & \cellcolor{cyan}0.8 & \cellcolor{green}0.7 & \cellcolor{cyan}0.8 & \cellcolor{cyan}0.8 & \cellcolor{green}0.7 & \cellcolor{cyan}0.8 & \cellcolor{green}0.7 & \cellcolor{cyan}0.8 & \cellcolor{green}0.6 & \cellcolor{cyan}0.8 & \cellcolor{cyan}0.8 & \cellcolor{green}0.6 & \cellcolor{cyan}0.8 & \cellcolor{cyan}0.8 & \cellcolor{cyan}0.9 & \cellcolor{green}0.7 & \cellcolor{green}0.7 & \cellcolor{green}0.7 & \cellcolor{green}0.7 & \cellcolor{green}0.7 & \cellcolor{green}0.7 & \cellcolor{green}0.7\\
        \hline 
        \end{tabular}%
    }
    \label{tab:cross_ptp}
\end{table*}

\begin{table*}
    \caption{Cross-performance table for neural network $f_{\rm recov}|P_{\rm a}>0.75$ for NNMs trained on selected Auriga galaxies (columns) evaluating the star particles from the same set of galaxies (rows). Cells are coloured based on performance thresholds, with magenta indicating a $f_{\rm recov} < 0.5$, green a $f_{\rm recov} < 0.75$, and blue a $f_{\rm recov} > 0.75$.}
    \resizebox{\textwidth}{!}{%
    \begin{tabular}{llllllllllllllllllllllllll}
    \hline
    Validation\textbackslash{}Training & 2 & 3 & 4 & 5 & 6 & 7 & 9 & 10 & 12 & 13 & 14 & 15 & 16 & 17 & 18 & 19 & 20 & 21 & 22 & 23 & 24 & 26 & 27 & 28 \\
    \hline
    2 & \cellcolor{magenta}0.4 & \cellcolor{green}0.7 & \cellcolor{magenta}0.5 & \cellcolor{green}0.6 & \cellcolor{green}0.5 & \cellcolor{magenta}0.1 & \cellcolor{green}0.7 & \cellcolor{magenta}0.4 & \cellcolor{magenta}0.4 & \cellcolor{magenta}0.2 & \cellcolor{green}0.6 & \cellcolor{magenta}0.3 & \cellcolor{magenta}0.5 & \cellcolor{green}0.6 & \cellcolor{green}0.6 & \cellcolor{magenta}0.4 & \cellcolor{magenta}0.1 & \cellcolor{green}0.5 & \cellcolor{cyan}0.8 & \cellcolor{magenta}0.5 & \cellcolor{green}0.6 & \cellcolor{green}0.7 & \cellcolor{green}0.6 & \cellcolor{green}0.7 \\
    3 & \cellcolor{magenta}0.2 & \cellcolor{magenta}0.3 & \cellcolor{magenta}0.3 & \cellcolor{magenta}0.3 & \cellcolor{magenta}0.2 & \cellcolor{magenta}0.0 & \cellcolor{magenta}0.3 & \cellcolor{magenta}0.1 & \cellcolor{magenta}0.1 & \cellcolor{magenta}0.0 & \cellcolor{magenta}0.3 & \cellcolor{magenta}0.2 & \cellcolor{magenta}0.3 & \cellcolor{magenta}0.3 & \cellcolor{magenta}0.3 & \cellcolor{magenta}0.2 & \cellcolor{magenta}0.0 & \cellcolor{magenta}0.3 & \cellcolor{magenta}0.3 & \cellcolor{magenta}0.2 & \cellcolor{magenta}0.3 & \cellcolor{magenta}0.3 & \cellcolor{magenta}0.3 & \cellcolor{magenta}0.4 \\
    4 & \cellcolor{green}0.5 & \cellcolor{green}0.6 & \cellcolor{magenta}0.3 & \cellcolor{magenta}0.4 & \cellcolor{magenta}0.4 & \cellcolor{magenta}0.1 & \cellcolor{magenta}0.4 & \cellcolor{magenta}0.4 & \cellcolor{magenta}0.4 & \cellcolor{magenta}0.2 & \cellcolor{green}0.6 & \cellcolor{magenta}0.2 & \cellcolor{magenta}0.4 & \cellcolor{green}0.5 & \cellcolor{magenta}0.4 & \cellcolor{magenta}0.3 & \cellcolor{magenta}0.2 & \cellcolor{green}0.6 & \cellcolor{green}0.5 & \cellcolor{magenta}0.5 & \cellcolor{magenta}0.4 & \cellcolor{green}0.7 & \cellcolor{green}0.6 & \cellcolor{magenta}0.5 \\
    5 & \cellcolor{green}0.6 & \cellcolor{cyan}0.9 & \cellcolor{green}0.6 & \cellcolor{cyan}0.8 & \cellcolor{green}0.7 & \cellcolor{magenta}0.1 & \cellcolor{cyan}0.8 & \cellcolor{green}0.6 & \cellcolor{green}0.6 & \cellcolor{magenta}0.3 & \cellcolor{cyan}0.8 & \cellcolor{magenta}0.4 & \cellcolor{green}0.6 & \cellcolor{green}0.7 & \cellcolor{cyan}0.8 & \cellcolor{magenta}0.5 & \cellcolor{magenta}0.2 & \cellcolor{cyan}0.8 & \cellcolor{cyan}0.9 & \cellcolor{green}0.7 & \cellcolor{green}0.7 & \cellcolor{cyan}0.8 & \cellcolor{cyan}0.8 & \cellcolor{cyan}0.8 \\
    6 & \cellcolor{green}0.6 & \cellcolor{cyan}0.8 & \cellcolor{green}0.7 & \cellcolor{cyan}0.8 & \cellcolor{green}0.6 & \cellcolor{magenta}0.1 & \cellcolor{cyan}0.8 & \cellcolor{green}0.6 & \cellcolor{green}0.6 & \cellcolor{magenta}0.3 & \cellcolor{green}0.7 & \cellcolor{green}0.5 & \cellcolor{green}0.7 & \cellcolor{cyan}0.8 & \cellcolor{cyan}0.8 & \cellcolor{green}0.6 & \cellcolor{magenta}0.3 & \cellcolor{green}0.7 & \cellcolor{cyan}0.8 & \cellcolor{green}0.6 & \cellcolor{cyan}0.8 & \cellcolor{green}0.7 & \cellcolor{cyan}0.8 & \cellcolor{cyan}0.8 \\
    7 & \cellcolor{cyan}0.8 & \cellcolor{cyan}0.9 & \cellcolor{green}0.6 & \cellcolor{cyan}0.8 & \cellcolor{cyan}0.8 & \cellcolor{magenta}0.3 & \cellcolor{green}0.7 & \cellcolor{green}0.7 & \cellcolor{green}0.7 & \cellcolor{green}0.5 & \cellcolor{cyan}0.8 & \cellcolor{magenta}0.4 & \cellcolor{cyan}0.8 & \cellcolor{cyan}0.8 & \cellcolor{green}0.7 & \cellcolor{magenta}0.5 & \cellcolor{magenta}0.4 & \cellcolor{cyan}0.8 & \cellcolor{cyan}0.9 & \cellcolor{green}0.7 & \cellcolor{green}0.7 & \cellcolor{cyan}0.8 & \cellcolor{cyan}0.8 & \cellcolor{cyan}0.8 \\
    9 & \cellcolor{green}0.7 & \cellcolor{cyan}0.8 & \cellcolor{green}0.7 & \cellcolor{green}0.7 & \cellcolor{green}0.6 & \cellcolor{magenta}0.1 & \cellcolor{cyan}0.8 & \cellcolor{magenta}0.5 & \cellcolor{green}0.6 & \cellcolor{magenta}0.2 & \cellcolor{green}0.7 & \cellcolor{green}0.5 & \cellcolor{green}0.7 & \cellcolor{green}0.7 & \cellcolor{cyan}0.8 & \cellcolor{green}0.7 & \cellcolor{magenta}0.2 & \cellcolor{green}0.7 & \cellcolor{cyan}0.8 & \cellcolor{green}0.5 & \cellcolor{cyan}0.8 & \cellcolor{green}0.7 & \cellcolor{green}0.7 & \cellcolor{cyan}0.8 \\
    10 & \cellcolor{cyan}1.0 & \cellcolor{cyan}1.0 & \cellcolor{cyan}1.0 & \cellcolor{cyan}1.0 & \cellcolor{cyan}1.0 & \cellcolor{magenta}0.4 & \cellcolor{cyan}1.0 & \cellcolor{cyan}0.9 & \cellcolor{cyan}0.9 & \cellcolor{green}0.7 & \cellcolor{cyan}1.0 & \cellcolor{cyan}0.8 & \cellcolor{cyan}1.0 & \cellcolor{cyan}1.0 & \cellcolor{cyan}1.0 & \cellcolor{cyan}0.9 & \cellcolor{green}0.7 & \cellcolor{cyan}1.0 & \cellcolor{cyan}1.0 & \cellcolor{cyan}0.9 & \cellcolor{cyan}1.0 & \cellcolor{cyan}1.0 & \cellcolor{cyan}1.0 & \cellcolor{cyan}1.0 \\
    12 & \cellcolor{magenta}0.5 & \cellcolor{green}0.7 & \cellcolor{magenta}0.4 & \cellcolor{green}0.6 & \cellcolor{magenta}0.5 & \cellcolor{magenta}0.1 & \cellcolor{green}0.6 & \cellcolor{magenta}0.4 & \cellcolor{magenta}0.4 & \cellcolor{magenta}0.2 & \cellcolor{green}0.6 & \cellcolor{magenta}0.3 & \cellcolor{green}0.5 & \cellcolor{green}0.6 & \cellcolor{green}0.6 & \cellcolor{magenta}0.3 & \cellcolor{magenta}0.1 & \cellcolor{green}0.6 & \cellcolor{cyan}0.8 & \cellcolor{magenta}0.4 & \cellcolor{green}0.6 & \cellcolor{green}0.6 & \cellcolor{green}0.6 & \cellcolor{green}0.6 \\
    13 & \cellcolor{green}0.7 & \cellcolor{cyan}1.0 & \cellcolor{green}0.6 & \cellcolor{cyan}0.9 & \cellcolor{cyan}0.9 & \cellcolor{magenta}0.1 & \cellcolor{cyan}0.8 & \cellcolor{cyan}0.8 & \cellcolor{green}0.7 & \cellcolor{magenta}0.4 & \cellcolor{cyan}0.9 & \cellcolor{magenta}0.4 & \cellcolor{green}0.7 & \cellcolor{cyan}0.8 & \cellcolor{cyan}0.8 & \cellcolor{green}0.5 & \cellcolor{magenta}0.3 & \cellcolor{cyan}0.9 & \cellcolor{cyan}1.0 & \cellcolor{cyan}0.8 & \cellcolor{cyan}0.8 & \cellcolor{cyan}0.9 & \cellcolor{cyan}0.9 & \cellcolor{cyan}0.9 \\
    14 & \cellcolor{green}0.6 & \cellcolor{cyan}0.9 & \cellcolor{magenta}0.5 & \cellcolor{green}0.7 & \cellcolor{green}0.7 & \cellcolor{magenta}0.1 & \cellcolor{green}0.7 & \cellcolor{green}0.6 & \cellcolor{green}0.5 & \cellcolor{magenta}0.2 & \cellcolor{green}0.7 & \cellcolor{magenta}0.3 & \cellcolor{green}0.6 & \cellcolor{green}0.7 & \cellcolor{green}0.7 & \cellcolor{magenta}0.4 & \cellcolor{magenta}0.2 & \cellcolor{cyan}0.8 & \cellcolor{cyan}0.8 & \cellcolor{green}0.6 & \cellcolor{green}0.6 & \cellcolor{cyan}0.8 & \cellcolor{cyan}0.8 & \cellcolor{green}0.7 \\
    15 & \cellcolor{cyan}0.9 & \cellcolor{cyan}1.0 & \cellcolor{cyan}0.9 & \cellcolor{cyan}1.0 & \cellcolor{cyan}1.0 & \cellcolor{magenta}0.3 & \cellcolor{cyan}1.0 & \cellcolor{cyan}1.0 & \cellcolor{cyan}0.9 & \cellcolor{green}0.7 & \cellcolor{cyan}1.0 & \cellcolor{green}0.6 & \cellcolor{cyan}0.9 & \cellcolor{cyan}0.9 & \cellcolor{cyan}1.0 & \cellcolor{green}0.7 & \cellcolor{green}0.6 & \cellcolor{cyan}1.0 & \cellcolor{cyan}1.0 & \cellcolor{cyan}0.9 & \cellcolor{cyan}0.9 & \cellcolor{cyan}1.0 & \cellcolor{cyan}1.0 & \cellcolor{cyan}1.0 \\
    16 & \cellcolor{green}0.7 & \cellcolor{cyan}1.0 & \cellcolor{green}0.7 & \cellcolor{cyan}0.9 & \cellcolor{cyan}0.8 & \cellcolor{magenta}0.1 & \cellcolor{cyan}0.9 & \cellcolor{green}0.7 & \cellcolor{green}0.7 & \cellcolor{magenta}0.3 & \cellcolor{cyan}0.8 & \cellcolor{magenta}0.4 & \cellcolor{cyan}0.8 & \cellcolor{cyan}0.8 & \cellcolor{cyan}0.9 & \cellcolor{green}0.6 & \cellcolor{magenta}0.3 & \cellcolor{cyan}0.8 & \cellcolor{cyan}1.0 & \cellcolor{green}0.7 & \cellcolor{cyan}0.8 & \cellcolor{cyan}0.8 & \cellcolor{cyan}0.9 & \cellcolor{cyan}0.9 \\
    17 & \cellcolor{cyan}0.8 & \cellcolor{cyan}1.0 & \cellcolor{cyan}0.9 & \cellcolor{cyan}0.9 & \cellcolor{cyan}0.8 & \cellcolor{magenta}0.1 & \cellcolor{cyan}0.9 & \cellcolor{green}0.7 & \cellcolor{green}0.7 & \cellcolor{magenta}0.3 & \cellcolor{cyan}0.8 & \cellcolor{green}0.7 & \cellcolor{cyan}0.9 & \cellcolor{cyan}0.9 & \cellcolor{cyan}0.9 & \cellcolor{cyan}0.8 & \cellcolor{magenta}0.3 & \cellcolor{cyan}0.9 & \cellcolor{cyan}1.0 & \cellcolor{green}0.7 & \cellcolor{cyan}0.9 & \cellcolor{cyan}0.8 & \cellcolor{cyan}0.9 & \cellcolor{cyan}0.9 \\
    18 & \cellcolor{cyan}0.8 & \cellcolor{cyan}0.9 & \cellcolor{cyan}0.9 & \cellcolor{cyan}0.9 & \cellcolor{green}0.7 & \cellcolor{magenta}0.1 & \cellcolor{cyan}0.9 & \cellcolor{green}0.7 & \cellcolor{green}0.7 & \cellcolor{magenta}0.3 & \cellcolor{cyan}0.8 & \cellcolor{green}0.6 & \cellcolor{cyan}0.9 & \cellcolor{cyan}0.9 & \cellcolor{cyan}1.0 & \cellcolor{cyan}0.8 & \cellcolor{magenta}0.3 & \cellcolor{cyan}0.8 & \cellcolor{cyan}1.0 & \cellcolor{green}0.7 & \cellcolor{cyan}0.9 & \cellcolor{cyan}0.8 & \cellcolor{cyan}0.9 & \cellcolor{cyan}0.9 \\
    19 & \cellcolor{green}0.7 & \cellcolor{cyan}0.9 & \cellcolor{magenta}0.5 & \cellcolor{green}0.7 & \cellcolor{cyan}0.8 & \cellcolor{magenta}0.2 & \cellcolor{green}0.7 & \cellcolor{green}0.7 & \cellcolor{green}0.7 & \cellcolor{magenta}0.4 & \cellcolor{cyan}0.8 & \cellcolor{magenta}0.3 & \cellcolor{green}0.6 & \cellcolor{green}0.7 & \cellcolor{green}0.7 & \cellcolor{magenta}0.4 & \cellcolor{magenta}0.3 & \cellcolor{cyan}0.8 & \cellcolor{cyan}0.9 & \cellcolor{cyan}0.8 & \cellcolor{green}0.6 & \cellcolor{cyan}0.9 & \cellcolor{cyan}0.9 & \cellcolor{cyan}0.8 \\
    20 & \cellcolor{green}0.6 & \cellcolor{green}0.6 & \cellcolor{magenta}0.3 & \cellcolor{magenta}0.4 & \cellcolor{magenta}0.4 & \cellcolor{magenta}0.1 & \cellcolor{magenta}0.4 & \cellcolor{magenta}0.3 & \cellcolor{magenta}0.4 & \cellcolor{magenta}0.2 & \cellcolor{green}0.6 & \cellcolor{magenta}0.2 & \cellcolor{magenta}0.3 & \cellcolor{green}0.5 & \cellcolor{magenta}0.3 & \cellcolor{magenta}0.2 & \cellcolor{magenta}0.1 & \cellcolor{green}0.6 & \cellcolor{green}0.6 & \cellcolor{magenta}0.5 & \cellcolor{magenta}0.4 & \cellcolor{cyan}0.8 & \cellcolor{green}0.6 & \cellcolor{magenta}0.4 \\
    21 & \cellcolor{magenta}0.4 & \cellcolor{green}0.6 & \cellcolor{magenta}0.4 & \cellcolor{green}0.6 & \cellcolor{magenta}0.5 & \cellcolor{magenta}0.1 & \cellcolor{green}0.6 & \cellcolor{magenta}0.4 & \cellcolor{magenta}0.4 & \cellcolor{magenta}0.2 & \cellcolor{green}0.6 & \cellcolor{magenta}0.3 & \cellcolor{magenta}0.5 & \cellcolor{green}0.6 & \cellcolor{green}0.6 & \cellcolor{magenta}0.4 & \cellcolor{magenta}0.2 & \cellcolor{magenta}0.5 & \cellcolor{green}0.6 & \cellcolor{magenta}0.4 & \cellcolor{green}0.6 & \cellcolor{green}0.5 & \cellcolor{green}0.5 & \cellcolor{green}0.6 \\
    22 & \cellcolor{cyan}0.9 & \cellcolor{cyan}1.0 & \cellcolor{cyan}0.9 & \cellcolor{cyan}0.9 & \cellcolor{green}0.7 & \cellcolor{magenta}0.2 & \cellcolor{cyan}0.9 & \cellcolor{green}0.7 & \cellcolor{green}0.7 & \cellcolor{magenta}0.3 & \cellcolor{cyan}0.9 & \cellcolor{green}0.7 & \cellcolor{cyan}0.9 & \cellcolor{cyan}0.9 & \cellcolor{cyan}1.0 & \cellcolor{cyan}0.9 & \cellcolor{magenta}0.3 & \cellcolor{cyan}0.9 & \cellcolor{cyan}0.9 & \cellcolor{green}0.7 & \cellcolor{cyan}0.9 & \cellcolor{cyan}0.8 & \cellcolor{cyan}0.9 & \cellcolor{cyan}0.9 \\
    23 & \cellcolor{green}0.6 & \cellcolor{green}0.7 & \cellcolor{green}0.7 & \cellcolor{green}0.7 & \cellcolor{green}0.5 & \cellcolor{magenta}0.1 & \cellcolor{green}0.7 & \cellcolor{magenta}0.5 & \cellcolor{green}0.5 & \cellcolor{magenta}0.2 & \cellcolor{green}0.6 & \cellcolor{magenta}0.4 & \cellcolor{green}0.7 & \cellcolor{green}0.7 & \cellcolor{green}0.7 & \cellcolor{green}0.5 & \cellcolor{magenta}0.2 & \cellcolor{green}0.7 & \cellcolor{green}0.7 & \cellcolor{magenta}0.5 & \cellcolor{green}0.7 & \cellcolor{green}0.6 & \cellcolor{green}0.7 & \cellcolor{green}0.7 \\
    24 & \cellcolor{green}0.6 & \cellcolor{cyan}0.9 & \cellcolor{green}0.6 & \cellcolor{cyan}0.8 & \cellcolor{green}0.6 & \cellcolor{magenta}0.1 & \cellcolor{green}0.7 & \cellcolor{green}0.6 & \cellcolor{green}0.6 & \cellcolor{magenta}0.2 & \cellcolor{green}0.7 & \cellcolor{magenta}0.4 & \cellcolor{green}0.7 & \cellcolor{green}0.7 & \cellcolor{cyan}0.8 & \cellcolor{magenta}0.5 & \cellcolor{magenta}0.2 & \cellcolor{green}0.7 & \cellcolor{cyan}0.9 & \cellcolor{green}0.6 & \cellcolor{green}0.7 & \cellcolor{green}0.7 & \cellcolor{cyan}0.8 & \cellcolor{cyan}0.8 \\
    26 & \cellcolor{magenta}0.4 & \cellcolor{green}0.5 & \cellcolor{magenta}0.5 & \cellcolor{magenta}0.5 & \cellcolor{magenta}0.3 & \cellcolor{magenta}0.0 & \cellcolor{green}0.5 & \cellcolor{magenta}0.3 & \cellcolor{magenta}0.3 & \cellcolor{magenta}0.1 & \cellcolor{magenta}0.5 & \cellcolor{magenta}0.3 & \cellcolor{magenta}0.5 & \cellcolor{magenta}0.5 & \cellcolor{magenta}0.5 & \cellcolor{magenta}0.4 & \cellcolor{magenta}0.1 & \cellcolor{magenta}0.5 & \cellcolor{magenta}0.5 & \cellcolor{magenta}0.3 & \cellcolor{magenta}0.5 & \cellcolor{magenta}0.4 & \cellcolor{magenta}0.5 & \cellcolor{green}0.5 \\
    27 & \cellcolor{magenta}0.5 & \cellcolor{green}0.7 & \cellcolor{green}0.6 & \cellcolor{green}0.6 & \cellcolor{magenta}0.3 & \cellcolor{magenta}0.1 & \cellcolor{green}0.7 & \cellcolor{magenta}0.3 & \cellcolor{magenta}0.4 & \cellcolor{magenta}0.1 & \cellcolor{green}0.5 & \cellcolor{magenta}0.4 & \cellcolor{green}0.6 & \cellcolor{green}0.6 & \cellcolor{green}0.7 & \cellcolor{magenta}0.5 & \cellcolor{magenta}0.1 & \cellcolor{green}0.5 & \cellcolor{green}0.7 & \cellcolor{magenta}0.3 & \cellcolor{green}0.6 & \cellcolor{magenta}0.5 & \cellcolor{green}0.6 & \cellcolor{green}0.6 \\
    28 & \cellcolor{magenta}0.3 & \cellcolor{magenta}0.4 & \cellcolor{magenta}0.3 & \cellcolor{magenta}0.4 & \cellcolor{magenta}0.2 & \cellcolor{magenta}0.1 & \cellcolor{magenta}0.4 & \cellcolor{magenta}0.2 & \cellcolor{magenta}0.3 & \cellcolor{magenta}0.1 & \cellcolor{magenta}0.4 & \cellcolor{magenta}0.2 & \cellcolor{magenta}0.3 & \cellcolor{magenta}0.5 & \cellcolor{magenta}0.3 & \cellcolor{magenta}0.3 & \cellcolor{magenta}0.1 & \cellcolor{magenta}0.3 & \cellcolor{magenta}0.4 & \cellcolor{magenta}0.2 & \cellcolor{magenta}0.4 & \cellcolor{magenta}0.4 & \cellcolor{magenta}0.4 & \cellcolor{magenta}0.4\\
    \hline 
    \end{tabular}%
    }
    \label{tab:cross_frec}
\end{table*}

\subsection{Conglomerated training Set}

One approach we can use to ameliorate issues that arise due to galactic satellite environments that do not well or fully describe the accreted stellar material in the disc is conglomeration. By constructing a training set out of a wide survey of satellite galaxy, and halo star particles we can attempt to train a single model that can perform well on a wide assortment of simulated galaxies. Table \ref{tab:networkperf_conglomerate} displays the results of training a NNM on the union of all the individual Auriga training sets, over a maximum of 100 training epochs, during which NNM weights were adjusted to model the training data. We can see that the variation of the results between galaxies is significantly reduced with P(TP) above 0.5 in all but 8 galaxies at a $P_{\rm a}$ threshold of 0.5, and none at higher thresholds. Recovery fraction varies more predictably with thresholds in $P_{\rm a}$, since the dependence on the environment-specific features is reduced. 

\begin{table*}
    \caption{Neural network results for a single NNM trained with data from every Auriga galactic environment.}
    \begin{tabular}{lllllll}
    \hline
    Halo & $f_{\rm recov} | P_{\rm a} > 0.9$ & $f_{\rm recov} | P_{\rm a} > 0.75$ & $f_{\rm recov} | P_{\rm a} > 0.5$ & $P(TP) | P_{\rm a} > 0.9$ & $P(TP) | P_{\rm a} > 0.75$ & $P(TP) | P_{\rm a} > 0.5$ \\
    \hline
        2 & 0.15 & 0.49 & 0.73 & 0.72 & 0.62 & 0.54 \\
        3 & 0.11 & 0.37 & 0.55 & 0.71 & 0.62 & 0.56 \\
        4 & 0.14 & 0.38 & 0.54 & 0.70 & 0.63 & 0.56 \\
        5 & 0.16 & 0.45 & 0.62 & 0.69 & 0.61 & 0.54 \\
        6 & 0.20 & 0.50 & 0.66 & 0.70 & 0.61 & 0.53 \\
        7 & 0.23 & 0.54 & 0.69 & 0.73 & 0.65 & 0.56 \\
        9 & 0.25 & 0.56 & 0.72 & 0.73 & 0.63 & 0.54 \\
        10 & 0.32 & 0.61 & 0.75 & 0.72 & 0.58 & 0.50 \\
        12 & 0.30 & 0.60 & 0.75 & 0.72 & 0.59 & 0.50 \\
        13 & 0.30 & 0.61 & 0.77 & 0.72 & 0.59 & 0.50 \\
        14 & 0.29 & 0.61 & 0.77 & 0.73 & 0.60 & 0.52 \\
        15 & 0.31 & 0.64 & 0.79 & 0.72 & 0.60 & 0.51 \\
        16 & 0.31 & 0.65 & 0.80 & 0.72 & 0.59 & 0.50 \\
        17 & 0.32 & 0.67 & 0.81 & 0.70 & 0.57 & 0.48 \\
        18 & 0.33 & 0.68 & 0.82 & 0.70 & 0.56 & 0.47 \\
        19 & 0.33 & 0.67 & 0.82 & 0.70 & 0.57 & 0.48 \\
        20 & 0.31 & 0.65 & 0.80 & 0.71 & 0.58 & 0.49 \\
        21 & 0.31 & 0.64 & 0.79 & 0.72 & 0.58 & 0.49 \\
        22 & 0.32 & 0.65 & 0.80 & 0.70 & 0.57 & 0.48 \\
        23 & 0.32 & 0.65 & 0.80 & 0.71 & 0.57 & 0.49 \\
        24 & 0.31 & 0.65 & 0.80 & 0.70 & 0.57 & 0.49 \\
        26 & 0.30 & 0.64 & 0.79 & 0.71 & 0.58 & 0.50 \\
        27 & 0.30 & 0.64 & 0.79 & 0.71 & 0.58 & 0.50 \\
        28 & 0.29 & 0.62 & 0.78 & 0.72 & 0.59 & 0.51 \\
    \hline
\end{tabular}
\label{tab:networkperf_conglomerate}
\end{table*}


\section{Conclusions}
\label{section6}

In this work we have developed a machine learning method {\maga} for identifying accreted stars purely with chemical abundances and age.  We  used a suite of simulations from the Auriga project to build the NNM framework and tested its performance. These galaxies have MW mass-sized halos with a variety of assembly histories.

We adopt two performance measures to quantify the classification performance in general, $P(TP)$ and $f_{\rm recov}$, calculated by assuming  a probability $P_a$ for a star particle to be classified as true-positive. 
Each performance indicator illustrates a different aspect of how reliable \maga~ labels are for each of our test galaxies. $P(TP)$ measures the precision of the labels, and $f_{\rm recov}$ measures the completeness of the detected stars.

They are also more reliable indicators of performance than simply accuracy, which is not tailored to specific cases, and is misleading in situations with heavily imbalanced populations (Fig.\ref{fig:ptp_v_frecov} and Fig.\ref{fig:median_f_FP}). We have  shown that, while \maga~ is based on chemical abundance information and ages, the classification results in stellar population which have the expected dynamical properties (Fig.\ref{fig:sat_source_comp}). For example, the True Negative stars are rotation dominated ($\epsilon_{\rm J} \sim 1$), radially extended and with low heights above the discs. Conversely, while True Positive stars are also dominated by rotation they show a larger dispersion in $\epsilon_{\rm J}$ which is agreement with the larger height achieved by this sample.  
Those stars misclassified have properties that distinguished themselves clearly (Fig.\ref{fig:Au14_epsj_rad}, \ref{fig:Au14_composite}, \ref{fig:Au22_composite} and \ref{fig:Au20_composite}).

Our NNM algorithm appears to identify endo-debris stars as in situ stars, as their chemical patterns  mirror those of the satellites from which the gas that formed them originated. These stars tend to be slightly less $\alpha$-enriched at given [Fe/H] \citep{tissera_2014}.
In general, because this gas component is transformed into stars as gas bound to the main host galaxy,
the new born stars are considered by the merger tree ground-truth to have been formed in-situ. This tension results in false-positives and lowers the calculated P(TP) values, however it does not imply that the \maga~ is not functioning as intended.These miss-classified stars can provide information about the accretion of gas-rich mergers by using kinematical and spatial information if available.
It is unknown how modelling of gaseous mixing will affect this, although by altering the chemical balance of the resulting stars to a point between that of the satellite and that of the primary galaxy, the $P_{\rm a}$ assigned by \maga~ will likely be lower, possibly reducing false-positive rates.
On the other hand, stellar material in the disc that originates from accreted satellites with comparable mass to the progenitor galaxy may prove difficult to disentangle with just this method. Additional, physical and kinematic measurements will likely be necessary in those cases as shown in Fig.\ref{fig:Au22_composite}.

We have also demonstrated the performance of our method on specific tasks, such as identification of GES-mass proxy stellar structures, and the completeness of the recovered accretion history for all sources of accreted stars (Table \ref{tab:enc_test}). While our method struggles to recover large portions of the accreted stellar population in situations where the satellite and halo training set is not a fair reflection of sources of the accreted stars in the disc region, in the majority of cases it is able to recover substantial fractions of them. We find that our training method generates NNM models that are generally able to recover pieces of nearly every accretion source that is present in the disc. This lends NNMs incredible power to aid in the deconstruction of galactic accretion histories, as shown in Fig.\ref{fig:h14_source_recovery}.

The best \maga~ performance is achieved when the most possible sources of accreted stars are included in the training data set (Fig.\ref{fig:sat_source_comp}). In cases where it is difficult to build a complete, accurate set of stars from satellites, we have shown that a selection of halo stars is sufficient for a training data set, at only a marginal cost to P(TP), or precision. A large training set which includes stars from a wide range of satellites will increase the performance of \maga. 
 When the assembly history of the stellar halos and the surviving satellites involved only small galaxies, there is a systematically smaller fraction of accreted stars in the discs, which makes it difficult to extract them.
Additionally, some accretion events could have contributed with large fraction of accreted stars to the discs but the remnants are also distributed close to the discs. As a consequence, these stars are excluded from the training sets to avoid confusion and, therefore, are  missed. 
These cases suggest that a training set including stars from the environments of other massive galaxies could be a way to identify these stars even when the current environments do not include massive satellites.


In fact, we find that training sets can be combined to great effect, and that the NNM algorithm trained on this conglomerated data can be generally applied and achieve more consistent results than \maga~ trained on a single galactic environment (Table \ref{tab:networkperf_conglomerate}). This further improves our method's applicability to observational data, by opening up the possibility of using stars and stellar populations from satellites and halos of other galaxies to broaden an investigation into the MW's stellar disc. This approach requires a large, varied training set, with multiple highly-resolved stellar discs and halos. This may become feasible as our observational techniques and technologies improve.

While our demonstrations use data from simulated MW-like galaxies, the input values can be gathered from any stellar data. In cases where we lack the required resolution to identify and accurately measure the chemical components of individual stars, chemical estimates from isochrones could be substituted, and final $P_{\rm a}$ estimates can be determined based on the likelihood of each potential set of isochrone parameters, and the output $P_{\rm a}$ of each particular configuration.
Hence, unlike previous efforts \citep{Ostdiek_2020}, which have shown success at applying neural networks to observational MW data, our method does not require external data for training.

In the future, \maga~ can be improved by inclusion of data from other simulation suites to determine which findings are general, and which are isolated to the Auriga simulations. Machine learning and neural networks are a powerful set of tools for large-scale astronomical classification problems. As we develop the proper language and measures to describe their performance in the unique setting of galaxy formation, we can begin to expect their flexibility to lead to further exploration of their applications in this field. 

\section*{Data Availability}
The data and programs underlying this article will be shared on reasonable request to the authors.

\section*{Acknowledgements}
TJT acknowledges support from the ANID PhD fellowship.
PBT acknowledges partial funding by Fondecyt 1200703/2020 (ANID) and Nucleo Milenio ERIS. PBT and FAG ANID Basal Project FB210003.FAG acknowledges support from ANID FONDECYT Regular 1211370 and from the Max Planck Society through a “Partner Group” grant. RG acknowledges financial support from the Spanish Ministry of Science and Innovation (MICINN) through the Spanish State Research Agency, under the Severo Ochoa Program 2020-2023 (CEX2019-000920-S). This project has been supported partially by the European Union Horizon 2020 Research and Innovation Programme under the Marie Sklodowska-Curie grant agreement No 734374.



\bibliographystyle{mnras}
\bibliography{biblio} 




\appendix
\section{Training Set Makeup Comparisons}
\begin{figure} 
    \includegraphics[width=\columnwidth]{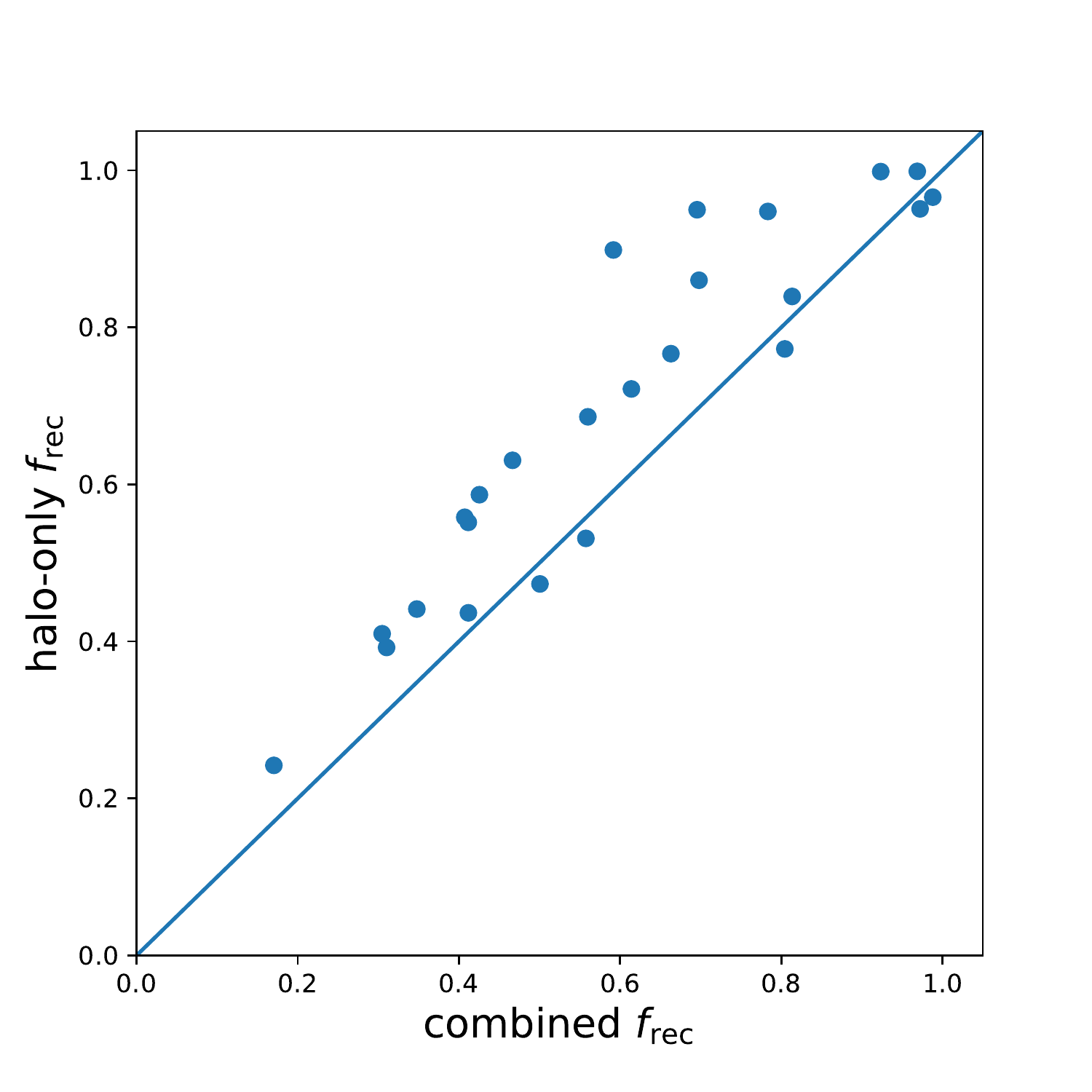}
    \includegraphics[width=\columnwidth]{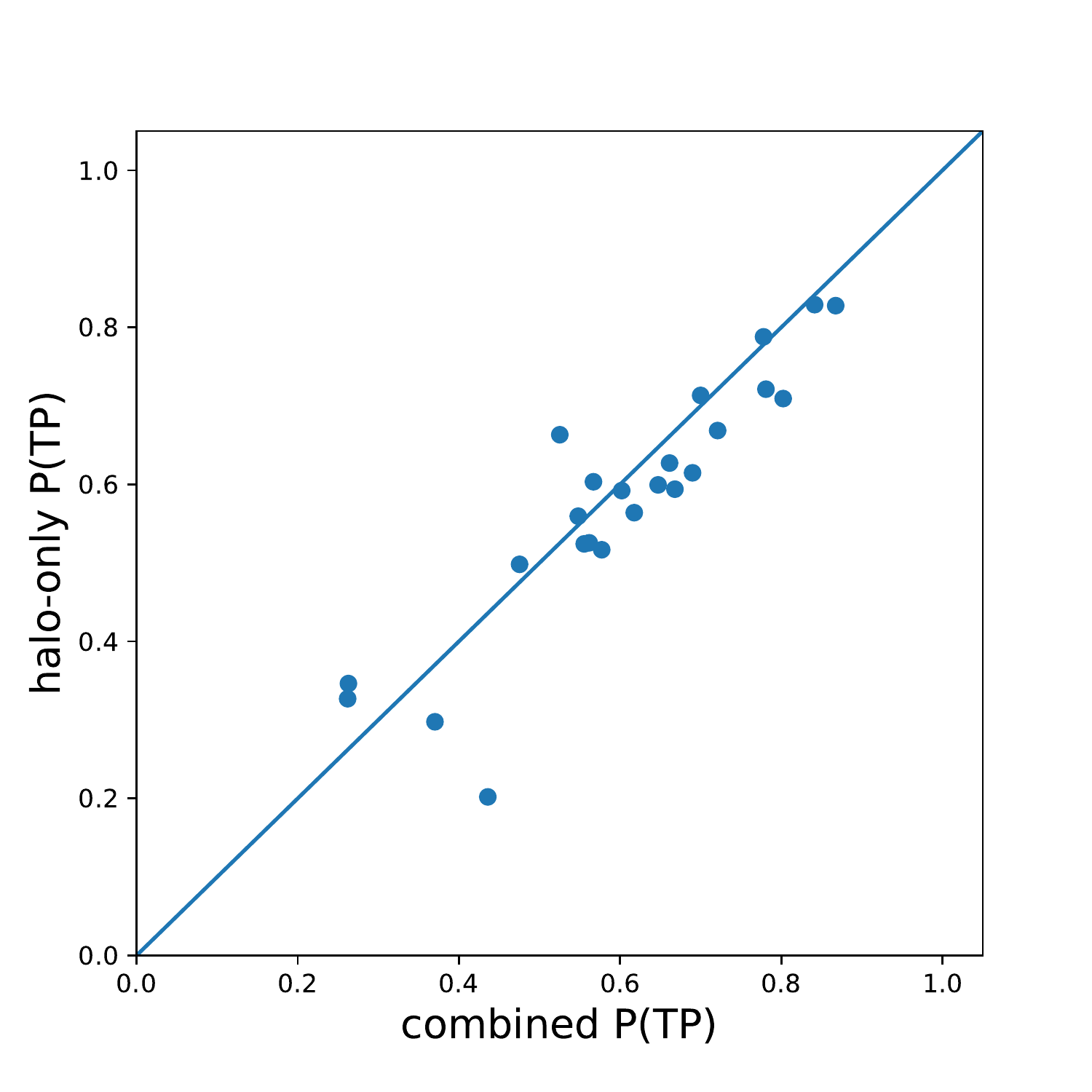}
    \caption{Comparison of NNM results when trained on the final combination data set, and a set composed entirely of halo stars greater than 25 kpc from the galactic center. At a threshold of $P_{\rm a} > 0.75$, halo-only NNMs have systematically higher $f_{\rm rec}$ and systematically lower P(TP), indicating that they are slightly less distinguishing than NNMs trained on data that includes satellite stars.}
    \label{fig:halo_comparison}
\end{figure}

While we have displayed the individual training sub-selection performances alongside each other in Figure \ref{fig:ptp_v_frecov}, we have also directly compared the best alternative training set composition with our final, combined choice in Figure \ref{fig:halo_comparison}, for a threshold of $P_{\rm a} > 0.75$. This second-best selection is entirely composed of stars from the galactic stellar halo past 25 kpc from the galactic center. On average, using this training set increases our $f_{\rm rec}$ with a mild decrease in P(TP). This can be explained by the fact that the stellar halo is almost entirely composed of stellar debris from satellites that have either passed through or merged with the central galaxy \citep{Bell_2008}. This implies that the chemistry-age trend information of any accreted stars in the galactic disc should be represented in this single set of training data. In cases where the halo is dominated by the debris from relatively few objects, however, extra care will be required to prevent the NNM from learning only to identify stars from these objects at the expense of others.

These results provide a good motivation for the use of halo-only training sets in cases where satellite stellar information is difficult to obtain, or of low quality. This will be an advantage particularly in an observational use-case, where significantly more information exists on individual halo stars than for stars in orbiting satellites.


\bsp	
\label{lastpage}
\end{document}